\documentclass[preprint,aps,prd,showpacs,nofootinbib,superscriptaddress,floatfix,
tightenlines]{revtex4-1}

\usepackage{amsmath}
\usepackage{graphicx}
\usepackage{subfigure}
\usepackage{epstopdf}
\usepackage{epsfig}
\usepackage{amssymb}
\usepackage{bm}
\usepackage{bbm}
\newcommand{\beq}{\begin{equation}}
\newcommand{\eeq}{\end{equation}}
\newcommand{\ov}{\overline}

\usepackage{color}

\begin{document}

\begin{flushright}
{\small KIAS-PREPRINT-P15006} 
\end{flushright}

\title{Higgs and Dark Matter Physics \\ in the Type-II Two-Higgs-Doublet Model
inspired by $E_6$ GUT}

\author{P. Ko}
\email[]{pko@kias.re.kr}
\affiliation{School of Physics, KIAS, Seoul 130-722, Korea}

\author{Yuji Omura}
\email[]{yujiomur@eken.phys.nagoya-u.ac.jp}
\affiliation{Department of Physics, Nagoya University, Nagoya 464-8602, Japan}

\author{Chaehyun Yu}
\email[]{chyu@kias.re.kr}
\affiliation{School of Physics, KIAS, Seoul 130-722, Korea}

\date{\today}

\begin{abstract}
\noindent
We study Higgs and dark matter physics in the type-II two-Higgs-doublet 
model (2HDM) with an extra $U(1)_H$ gauge symmetry,
inspired by the $E_6$ grand unified theory (GUT). From the viewpoint 
of the bottom-up approach,  the additional $U(1)_H$ gauge symmetry 
plays a crucial role in avoiding the tree-level flavor changing neutral 
currents mediated by neutral Higgs bosons in general 2HDMs.  
In the model with $U(1)_H$ gauge symmetry, which has Type-II Yukawa couplings,  
we have to introduce additional chiral fermions that are charged under 
the $U(1)_H$ gauge symmetry as well as under the Standard-Model (SM) 
gauge symmetry in order to cancel chiral gauge anomalies. 
For the $U(1)_H$ charge assignment and the extra matters, we adopt the ones 
inspired by the $E_6$ GUT: the extra quark-like and lepton-like fermions 
with the non-trivial $U(1)_H$  charges.  We discuss their contributions 
to the physical observables, such as the measurements of Higgs physics 
and electro-weak interactions, and investigate the consistency 
with the experimental results. 
Furthermore, we could find extra neutral particles like the SM neutrinos 
after the electro-weak symmetry breaking, and they could be stable, 
because of the remnant symmetry after $U(1)_H$ symmetry breaking.
We also discuss the thermal relic density and the (in)direct-detections 
of this dark matter candidate.

\end{abstract}

\maketitle

\section{Introduction}
\label{sec:introduction}

Adding extra Higgs doublets to the Standard Model (SM) would be one of the most 
attractive and the simplest ways to consider the extension of the SM.
In fact, such extra Higgs doublets are present in many Beyond Standard Models (BSMs)
motivated by some theoretical problems of the SM such as gauge hierarchy problem.  Theoretical and phenomenological aspects of multi-Higgs-Doublet models have been 
widely discussed so far.
Especially, two-Higgs-Doublet models (2HDM) with (softly broken) $Z_2$ Higgs symmetry
are well-investigated, 
motivated by supersymmetry, grand unification theories (GUT), Higgs and dark matter 
physics  (see Ref. \cite{Branco:2011iw} for recent reviews). 
Also a lot of interests on this model have been drawn in light of 
new LHC data~\cite{2HDMLHC}.

The softly broken discrete $Z_2$ Higgs symmetry is introduced to avoid the tree-level 
flavor changing neutral currents (FCNCs) \'{a} la  Natural Flavor Conservation 
(NFC) criterion \cite{Glashow}.   The resulting 2HDMs predict the so-called minimal flavor 
violation, where the FCNCs  mediated by neutral Higgs bosons are suppressed by 
the CKM matrix and thus phenomenologically safe.

In order to avoid too large tree-level FCNCs, the present authors made new proposals 
of  (flavor-dependent) gauged $U(1)_H$ Higgs symmetry instead of the $Z_2$ symmetry 
in Refs. \cite{Ko-2HDM,Ko-2HDMtype1,Ko-Top,Ko-Top2}.
In the 2HDMs with $U(1)_H$ symmetry (denoted as 2HDM$_{\rm U(1)}$ hereafter),
two Higgs doublets are charged under new local $U(1)_H$ gauge symmetry, and they 
break both electroweak (EW) and $U(1)_H$ Higgs gauge symmetries.  In this new proposal, 
SM fermions have to be charged under $U(1)_H$; otherwise one cannot write the realistic  Yukawa 
couplings at the renormalizable level. 

The 2HDM$_{\rm U(1)}$ is strongly constrained 
by the measurements of the electroweak precision observables (EWPOs), 
as well as the collider searches for $Z'$ and Higgs boson. 
In fact, the present authors investigated the constraints in the type-I 2HDM$_{U(1)}$ in 
detail,  and discussed the current status of the Type-I 2HDM$_{\rm U(1)}$ in light of the 
recent LHC results on Higgs properties and provided the future prospects in 
Ref.~\cite{Ko-2HDMtype1}.    Also they constructed the inert 2HDM model with $U(1)_H$ 
gauge symmetry and showed that the light dark matter (DM) mass region below 
$\sim m_W$ is widely open if the $Z_2$ symmetry is implemented  into local $U(1)_H$ 
gauge symmetry, due to newly open annihilation channels of the DM pair into
the extra $U(1)_H$ gauge boson(s):
$HH \rightarrow Z_H Z_H , Z Z_H$, which are not present in the ordinary inert 2HDM with 
discrete $Z_2$ symmetry \cite{Ko-IDM}. 
In fact this phenomenon is very generic in dark matter models with local dark gauge 
symmetries~\cite{Baek:2013qwa,Baek:2013dwa,Ko:2014nha,Ko:2014bka,Baek:2014kna,Ko:2014lsa}. 
Note that the $U(1)_H$ gauge symmetry is nothing but local dark gauge symmetry, 
since it acts only on the inert doublet, and not to the SM fields at all.

We may have to introduce extra chiral fermions to avoid gauge anomalies depending on 
the $U(1)_H$ charge assignments to the SM fermions. 
In Ref.  \cite{Ko-2HDMtype1}, it was shown that the anomaly-free $U(1)_H$ charge assignments to the SM fermions are possible in the Type-I 2HDM$_{\rm U(1)}$, 
so the fermion sector is just the same as the SM case
except right-handed neutrinos.
However, in other types of 2HDMs, $U(1)_H$ becomes anomalous without extra chiral fermions, and then we face the strong constraints  on extra fermions from various experiments. For example,  in the Type-II 
2HDM$_{U(1)}$, which is the main subject of this work, there is no solution for the  
anomaly-free conditions without extra chiral fermions, as discussed in Ref. \cite{Ko-2HDM}. 
The extra particles would be colored and carry the electric charges.
Hence, they would be 
produced and detected at the LEP and hadron colliders, depending on their masses 
~\cite{Ko-2HDM}.  Therefore such additional particles charged under $U(1)_H$ and/or 
the SM gauge groups would suffer from strong theoretical and experimental constraints. 
On the other hand, some of them might be stable (or long-lived enough) and could  be 
good cold dark matter (CDM) candidates as pointed out in  Refs.~\cite{Ko-2HDM, Ko-IDM}. 
Their stability could be guaranteed by the remnant symmetry of  $U(1)_H$ \cite{Ko-IDM}.  
 
In this paper, we study Higgs and dark matter physics, as well as the experimental and theoretical 
constraints, in the Type-II 2HDM$_{U(1)}$, inspired by $E_6$ GUT. 
From the point of view of the bottom-up approach, there are many choices for the 
$U(1)_H$ charge assignment to realize the Type-II Yukawa couplings,  where one Higgs 
doublet couples with the up-type fermions and the other one couples with the down-type 
fermions in accordance with NFC.  One well-known $U(1)_H$ Higgs gauge symmetry 
would be the one predicted by GUT, such as $E_6$ and $SO(10)$ GUT.  
The rank of $E_6$ gauge group is $``6"$, so that $E_6$ predicts $2$ extra $U(1)$  symmetries and 
the SM fermions, as well as  extra fermions to  make the models free of gauge anomalies.  
They could be derived from the three-family ${ \bf 27}$  representations at low energy, 
in the supersymmetric $E_6$ model.\footnote{The general   analysis of $Z'$ in the $E_6$ 
GUT has been done in Refs. \cite{E6Zprime1,E6Zprime}. 
One can also see the reviews of the $E_6$ GUT\cite{E6review,E6review2}.}
If we assume that $U(1)_H$ is originated from breaking of two $U(1)$ symmetries at low 
energy scale,  the $U(1)_H$ charges are  predicted  explicitly  by the RG flow and the 
decoupling scales of the extra fields \cite{Babu,Rizzo}.  
The representative $U(1)_H$ charge assignments are $U(1)_{\psi}$, $U(1)_{\chi}$, and $U(1)_{\eta}$,  
and they face stringent constraints from the Drell-Yan (DY) processes from 
hadron colliders~\cite{E6bound}.   However, if we assume $U(1)_H$ is the so-called 
leptophobic $U(1)_b$ under which the SM leptons are not charged~\cite{E6leptophobic,E6leptophobic2,E6leptophobic3,E6leptophobic4,Babu,Rizzo}, 
one could evade the strong constraints from the Drell-Yan processes, and the 
$U(1)_H$  gauge boson could be as low as $\sim O(100)$ GeV.  

In our type-II 2HDM$_{U(1)}$, we shall assign one gauged $U(1)_H$ symmetry,  which 
may be derived from the $E_6$ GUT model, assuming one of the two $U(1)$'s is broken 
at a high scale.  From the viewpoint of the top-down approach, the $U(1)_H$ may be 
fixed once we chose the broken $U(1)$ symmetry, because they may be approximately 
orthogonal each other.   For instance, $U(1)_H$ may be the linear combination of $U(1)_Y$ 
and $U(1)_{\psi}$ when $U(1)_{\chi}$  is broken at high energy.

In this paper, we define the $U(1)_H$ as the leptophobic $U(1)_b$ which is the linear combination of  
$U(1)_{\eta}$ and $U(1)_Y$ taking the bottom-up approach. This is  because we could expect that such  
leptophobic  interaction may be sizable enough so that we may be able to observe new physics effects at 
colliders and dark matter experiments as demonstrated in the following. Furthermore, we consider the 
Yukawa couplings which respect both $U(1)_b$ and  $U(1)_{\psi}$ (or $U(1)_{\chi}$), in order to avoid 
the FCNCs induced by the mass mixings between extra fermions and the SM fermions 
\footnote{ $U(1)_{\psi}$ and $U(1)_{\chi}$ are not orthogonal to the leptophobic $U(1)_b$.
We will also give comments on the case with $U(1)_H \equiv U(1)_{\psi}$ or $U(1)_{\chi}$.}. 
 
Besides, we introduce only two Higgs doublets and the minimal set of the chiral fermions for 
the anomaly-free conditions which could be coming from the three-family of fundamental ${\bf 27}$'s.   
The extra fermions consist of the quark-like and lepton-like particles whose  charges under  the SM gauge 
groups are the same as the right-handed down quarks, the left-handed leptons and the right-handed 
neutrinos.   After EW symmetry breaking, the extra leptons are decomposed into neutral and charged 
particles just like neutrinos and charged leptons in the SM. In fact, there are 9 extra neutral and 6 charged 
particles, as we will see in Sec. II, and we could find the lightest neutral particle among them. 
The extra $U(1)_H$ symmetries are spontaneously broken but the remnant $Z_2^{{\rm ex}}$ symmetry is 
conserved. 
The lightest particle is charged under $U(1)_H$ and has odd parity under the remnant $Z_2^{{\rm ex}}$ symmetry, 
so that it becomes a good dark matter candidate.  DM will interact with SM particles through $Z_H$ and 
scalar boson exchanges.

This paper is organized as follows.  In Sec. \ref{section2}, we introduce the setup of our 2HDM$_{U(1)}$, 
presenting the extra $U(1)_H$ charge assignments to the SM fermions  and extra chiral fermions 
for the anomaly cancellation.  Then we discuss the interactions of the extra particles and the stability of 
the CDM candidate in Sec. \ref{section3}.    Then, we study the contributions of the $U(1)_H$ gauge boson 
and the extra fermions to the EWPOs, Higgs signals and phenomenology of CDM in Sec. \ref{section4}, \ref{section5} and \ref{section6}, respectively. 
Finally, Sec. \ref{section6} is devoted to summary. 
The gauge interactions and the vacuum polarizations  in our models are 
introduced in Appendix~\ref{gaugeinteraction} and \ref{ewpos}.


\section{Type-II 2HDM with Higgs symmetry}
\label{section2}

In order to realize the minimal flavor violation,  one fermion sector should couple with one Higgs doublet 
\'{a} la NFC.   Such Yukawa couplings can be realized by assuming an additional symmetry that 
distinguishes the two Higgs doublets: $Z_2$ symmetry \cite{Glashow} or gauged $U(1)_H$ symmetry \cite{Ko-2HDM}.
In the 2HDM$_{U(1)}$, the SM particles are
also charged under the additional gauge symmetry and extra chiral fermions
might be required to cancel the anomaly. In Ref. \cite{Ko-2HDMtype1},
the type-I 2HDM$_{U(1)}$ is mainly discussed 
and the gauged $U(1)_H$ symmetry is anomaly-free without any extra chiral fermions
except right-handed neutrinos.
In the type-II 2HDM$_{U(1)}$, the anomaly-free conditions 
cannot be satisfied without extra fermions \cite{Ko-2HDM}, so that we have to 
consider the more complex matter content and $U(1)_H$ charge assignment.
We could consider many models where the gauge anomalies are canceled by the extra fields as discussed in Ref. \cite{Ko-2HDM}.
In this section, we introduce the type-II 2HDM$_{U(1)}$ inspired by $E_6$ GUT. 

\subsection{Type-II 2HDM with gauged $U(1)_H$ symmetry inspired by $E_6$ GUT}
\label{sec:E6 Model}

The scalar potential of general 2HDMs with $U(1)_H$ is completely fixed by local  gauge invariance and renormalizability: 
\begin{eqnarray}
V&=& \Hat{m}^2_1(|\Phi|^2) H^{\dagger}_1 H_1+ \Hat{m}^2_2(|\Phi|^2) H^{\dagger}_2 H_2 
+m^2_{\Phi} |\Phi|^2 + \lambda_{\Phi} |\Phi|^4 - \left ( m^2_{3} (\Phi) H^{\dagger}_1 H_2 
+h.c.  \right )  \nonumber \\
&&+ \frac{\lambda_1}{2} (H^{\dagger}_1 H_1)^2 + \frac{\lambda_2}{2} (H^{\dagger}_2 H_2)^2 
+ \lambda_{3} (H^{\dagger}_1 H_1)(H^{\dagger}_2 H_2)+ \lambda_{4} |H^{\dagger}_1 H_2|^2  . 
\label{eq:potential}
\end{eqnarray}
Here $\Phi$ is a SM singlet  complex scalar field with $U(1)_H$ charge, $q_{\Phi}$, and contributes to 
the $U(1)_H$ symmetry breaking.  $\Hat{m}^2_i(|\Phi|^2)$ $(i=1,2)$ and $m^2_{3} (\Phi)$ are functions of 
$\Phi$ only:
\[
\Hat{m}^2_i(|\Phi|^2)=m_i^2+ \widetilde{\lambda}_i |\Phi|^2 
\] 
at the renormalizable level.   The function $m^2_{3} (\Phi)$ is fixed by the $U(1)_H$ charges ($q_{H_i}$) of the Higgs doublets ($H_i$) and $q_{\Phi}$, and 
$m^2_{3} (\langle \Phi \rangle)=0$  is satisfied at $\langle \Phi \rangle=0$: $m^2_{3} (\Phi) = \mu \Phi^n$, 
with $n \equiv (q_{H_1}-q_{H_2})/q_{\Phi}$.  The parameter $\mu$ can be rendered real after suitable 
redefinition of the phase of $\Phi$.  Note that the $\lambda_5$ term 
\[
\frac{1}{2} \lambda_5 [(H_1^\dagger H_2)^2+h.c.]
\]
in usual 2HDMs does not appear in the potential of our model because we employ a continuous  $U(1)_H$ 
gauge symmetry rather than a discrete $Z_2$ symmetry. 

The Yukawa couplings in the Type-II 2HDMs are defined as 
\beq
\label{potential}%
V_y= y^U_{ij} \overline{Q_L}^i \widetilde{H}_2 U^j_R+ y^D_{ij} \overline{Q_L}^i H_1 D^j_R+ y^E_{ij} \overline{L}^i H_1 E^j_R+ y^N_{ij} \overline{L}^i \widetilde{H}_2 N^j_R+h.c..
\eeq
Note that the $H_1$ and $H_2$ should carry different $U(1)_H$ charges in order to distinguish these two. 
\footnote
{For $U(1)_\chi$, two Higgs doublets carry the same charges, but right-handed up-type and down-type quarks
have the different $U(1)_\chi$ charges. 
}
In the Type-II 2HDM inspired by $E_6$, the charge assignments 
for the SM particles and the Higgs doublets are given in Table~\ref{table1}.

\begin{center}
\begin{table} \label{table1}
\begin{tabular}{|c|c|c|c||c|c|c|c|c|}\hline
             & $SU(3)$ & $SU(2)$  & $U(1)_Y$ & $U(1)_b$ & $U(1)_{\psi}$ & $U(1)_{\chi}$ & $U(1)_{\eta}$    \\ \hline  
 $Q^i$  & $3$         &$2$      &         $1/6$              &    $-1/3$ & $1$  & $-1$ & $-2$       \\ \hline 
  $U^i_{R}$  & $3$         &$1$      &       $2/3$        &  $2/3$    & $-1$  & $1$ & $2$       \\ \hline 
   $D^i_R$  & $3$         &$1$      &         $-1/3$              &    $-1/3$  & $-1$  & $-3$ & $-1$       \\ \hline 
   $L_i$  & $1$         &$2$      &            $-1/2$        &      $0$  & $1$  & $3$ & $1$    \\ \hline 
  $E^i_R$  & $1$         &$1$      &           $-1$            &   $0$    & $-1$  & $1$ & $2$     \\ \hline 
   $N^i_R$  & $1$         &$1$      &           $0$            &    $1$   & $-1$  & $5$ & $5$     \\ \hline 
    $H_1$  & $1$         &$2$      &          $1/2$            &   $0$     & $2$  & $2$ & $-1$      \\ \hline    
   $H_2$  & $1$         &$2$      &          $1/2$            &   $1$      & $-2$  & $2$ & $4$     \\ \hline 
\end{tabular}
\caption{Charge assignments of the SM fermions under the SM gauge group and various $U(1)$ subgroups  
of $E_6$ group
}
\label{table1}
\end{table} 
\end{center}

Let us assume that $E_6$ gauge symmetry breaks down as
\beq\label{E6}
E_6 \rightarrow SO(10) \times U(1)_{\psi} \rightarrow  SU(5) \times U(1)_{\chi} \times U(1)_\psi .
\eeq
The linear combination of $U(1)_{\psi}$ and  $U(1)_{\chi}$ gives $U(1)_{\eta}$,
and the leptophobic $U(1)_b$ is defined by their linear combinations with $U(1)_Y$ \cite{E6leptophobic,E6leptophobic2,E6leptophobic3,E6leptophobic4,Babu,Rizzo}:
\begin{eqnarray}
Q_{\eta}&=&\frac{3}{4} Q_{\chi}-\frac{5}{4} Q_{\psi}, \\
Q_b&=&\frac{1}{5} (Q_{\eta}+2Q_Y).
\end{eqnarray}
We can see the charge assignment for each $U(1)$ symmetry in Table~\ref{table1}.

The U(1) charge assignments of the SM fermions do not satisfy the anomaly-free conditions, 
and we have to introduce the following extra chiral fermions for anomaly cancellation: 
\beq
q^i_{L}, ~q^i_{R},~l^i_{L},~l^i_{R},~n^i_L.
\eeq
Here $n^i_L$ is neutral, and  $(q^i_{L},q^i_{R})$ and $(l^i_{L},l^i_{R})$ are vector-like fermions under 
the SM gauge groups.  Their $U(1)$ charges are chiral, as shown in Table~\ref{table2}.
\begin{center}
\begin{table} 
\begin{tabular}{|c|c|c|c||c|c|c|c|c|}\hline
             & $SU(3)$ & $SU(2)$  & $U(1)_Y$ & $U(1)_b$ & $U(1)_{\psi}$ & $U(1)_{\chi}$ & $U(1)_{\eta}$    \\ \hline   
  $q^i_{L}$  & $3$         &$1$      &       $-1/3$        &  $2/3$    & $-2$  & $2$ & $4$       \\ \hline 
    $q^i_{R}$  & $3$         &$1$      &       $-1/3$        &  $-1/3$    & $2$  & $2$ & $-1$       \\ \hline 
   $l^i_L$  & $1$         &$2$      &            $-1/2$        &      $0$  & $-2$  & $-2$ & $1$    \\ \hline 
  $l^i_R$  & $1$         &$2$      &           $-1/2$            &   $-1$    & $2$  & $-2$ & $-4$     \\ \hline 
   $n^i_L$  & $1$         &$1$      &           $0$            &    $-1$   & $4$  & $0$ & $-5$     \\ \hline 
\end{tabular}
\caption{Charge assignments of the exotic chiral fermions under the SM gauge group and various 
$U(1)$ subgroups of $E_6$.
}
\label{table2}
\end{table} 
\end{center}
The generation index, $i$, corresponds to those of the SM fermions, and  anomaly-free conditions are 
achieved within each generation.  
In the $E_6$ GUT, the SM fermions and these extra chiral fermions  are nicely embedded into three-family  
$\bf{27}$ representations. 
 
We have also introduced one extra complex scalar, $\Phi$, which is a singlet under the SM gauge group,
in order to break $U(1)_H$ spontaneously and generate the mass terms of the extra fermions.
Let us define the charges of $\Phi$ as shown in Table~\ref{table3} \footnote{In the supersymmetric $E_6$ model, 
$\Phi^*$  could be interpreted as the superpartner of $n_L$.}. 
\begin{center}
\begin{table} 
\begin{tabular}{|c|c|c|c|c|c|c|c|c|}\hline
             & $SU(3)$ & $SU(2)$  & $U(1)_Y$ & $U(1)_b$ & $U(1)_{\psi}$ & $U(1)_{\chi}$ & $U(1)_{\eta}$    \\ \hline   
  $\Phi$  & $1$         &$1$      &       $0$        &  $1$    & $-4$  & $0$ & $5$       \\ \hline 
\end{tabular}
\caption{Charge assignments of a singlet scalar $\Phi$ under the SM gauge group and $U(1)$ subgroup 
of $E_6$. This scalar $\Phi$ makes an additional contribution to $U(1)$ symmetry breaking.
}
\label{table3}
\end{table}
\end{center}
Then the Yukawa couplings which respect all extra $U(1)$ symmetries are  given by 
\beq
\label{eq:lepton mass}
V^{\rm ex}_y= y^q_{ij} \Phi \ov{q_L}^i q_R^j +y^l_{ij} \Phi \ov{l_L}^i l_R^j + y^n_{ij} \ov{l_R}^i 
\widetilde{H}_1 n^j_L +y'^n_{ij} \ov{l^c_L}^i  H_2 n^j_L+h.c..
\eeq
When $\Phi$ develops a nonzero vacuum expectation value (VEV), 
$q^i_{L,R}$ and $l^i_{L,R}$ would become massive through the 
Yukawa couplings.

As we discussed in Sec.~\ref{sec:introduction}, we assume  only one $U(1)_H$ gauge symmetry, which is the linear combination of the all $U(1)$ symmetries
and remains at low energy, whereas the other $U(1)$ from $E_6$ is spontaneously 
broken at the high energy scale. If $U(1)_{\psi}$ or $U(1)_{\chi}$ is broken,
the following Yukawa couplings would be allowed,
\beq\label{eq:mass mixing}
V^{\rm FCNC}_y= c^D_{ij} \Phi \ov{q_L}^i D_R^j +c^L_{ij} \Phi \ov{L}^i l_R^j +c^Q_{ij} \ov{Q_L}^i q_R^j H_1
+c^E_{ij}  \ov{l_L}^i E_R^j H_1+c^N_{ij} \ov{l_L}^i N_R^j \widetilde{H}_2  + h.c.,
\eeq
and the extra charged fermions mix with the SM quarks, and charged leptons and tree-level FCNC 
interactions will appear in general.   Hence we simply assume that $U(1)_{\psi}$ or $U(1)_{\chi}$ is broken 
at some energy,  but  the remnant symmetry of $U(1)_{\psi}$ or $U(1)_{\chi}$ still holds down to low energy 
scale to suppress the FCNCs. In fact, $U(1)_{\psi}$ $(U(1)_{\chi})$ breaks to $Z^{\psi}_2$ $(Z^{\chi}_2)$, 
if only the VEVs of $H_{1,2}$ and $\Phi$ break the $U(1)$ symmetry. 
The SM fermions are even and the extra fermions are odd under the remnant $Z_2$ symmetry, 
so that the mass-mixing terms between the SM and the extra fermions in 
Eq.~(\ref{eq:mass mixing}) are forbidden. 
We could also consider the case that $U(1)_H$ is identical to $U(1)_{\psi}$, for instance, 
and then the Yukawa couplings in Eq. (\ref{eq:mass mixing}) could be forbidden. 
In the sections for phenomenology, we adopt the $U(1)_b$ as the $U(1)_H$,
and investigate the impact of the new interaction inspired by $E_6$ GUT,
so that we simply assume that the mass mixings are forbidden by 
$Z^{\psi}_2$ ($Z^{\chi}_2$) symmetry at that time.
We shall also give a comment on the $U(1)_H \equiv U(1)_{\psi}$ case.

\section{Stability of the extra particles and the dark matter candidate}
\label{section3}
In this section, we briefly summarize the mass spectrum of the extra chiral fermions and discuss 
their stability. 
\subsection{Extra Leptons} \label{extraleptons}
Additional chiral fermions, $l^i_I$ $(I=L,R)$ and $n^i_L$, are color-singlets and their SM
charge assignment is the same as the one of the SM leptons and right-handed neutrino.
After the EW symmetry breaking, a doublet $l^i_I$ would split into a charged and a neutral fermions 
like the SM left-handed lepton doublets. The charged fermions become massive due to the nonzero 
$\langle \Phi \rangle$,   and the masses of the neutral fermions and $n^i_L$ are given by 
$\langle \Phi \rangle$ and $\langle H_{1,2} \rangle$ following Eq. (\ref{eq:lepton mass}).
The mass matrix for $l_I^T=( \widetilde{\nu}_I,\widetilde{e}_I)^T$ and $n_L$ is 
\begin{eqnarray}
{\cal L}_{  \nu}&=&-\frac{1}{2}
\begin{pmatrix} \ov{\widetilde{\nu}^c_L} & \ov{ \widetilde{\nu}_R} &  \ov{ n^c_L} \end{pmatrix}
\begin{pmatrix} 0 & m_{\widetilde{e}} & m_M \\  m_{\widetilde{e}} & 0 & m_D  \\ m_M & m_D & 0  \end{pmatrix}
\begin{pmatrix} \widetilde{\nu}_L \\  \widetilde{\nu}^c_R \\ n_L  \end{pmatrix} +h.c.  \\
&=&-\frac{1}{2}\begin{pmatrix} \ov{N_1} & \ov{ N_2} &  \ov{ N_3} \end{pmatrix}
\begin{pmatrix} m_1 & 0 & 0 \\  0 & m_2 & 0  \\ 0 & 0 & m_3   \end{pmatrix}
\begin{pmatrix} N_1 \\  N_2 \\ N_3  \end{pmatrix}.
\end{eqnarray}
Each element is defined as $ m_{\widetilde{e}} =y^l v_\Phi/\sqrt{2}$,  $ m_{D} =y^n v \cos \beta/\sqrt{2}$, and  
$ m_{M} =y'^n v \sin \beta/\sqrt{2}$. $ m_{\widetilde{e}}$ is the mass of the charged fermion, $\widetilde{e}$.
In general, $m_{\widetilde{e}}$, $m_{D}$, and $m_M$ are $3\times 3$ matrices in flavor space, and
would not be diagonal. In the following, we simply assume that dimensionless constants 
in Eq. (\ref{eq:lepton mass}) are flavor-blind  and we omit the flavor index $i$ in  $m_{\widetilde{e}}$, 
$m_{D}$, and $m_M$.  
At present, studying more general cases would be beyond the scope of this paper,
lacking any direct evidence of new particles at the LHC.

When $(m_{\widetilde{e}}^2+m_D^2+m_M^2)^3-27 (m_{\widetilde{e}}m_Dm_M)^2 \geq 0$ is satisfied, the mass eigenvalues, $m_1$, $m_2$ and $m_3$, are given by
\beq
2\sqrt{\frac{m_{\widetilde{e}}^2+m_D^2+m_M^2}{3}} \cos \frac{ \theta}{3},~2\sqrt{\frac{m_{\widetilde{e}}^2+m_D^2+m_M^2}{3}} \cos \left ( \frac{ \theta}{3} \pm \frac{2\pi}{3} \right ),
\eeq
where $\theta$ is defined as
\beq
\tan \theta= \frac{\sqrt{\{(m_{\widetilde{e}}^2+m_D^2+m_M^2)/3\}^3- (m_{\widetilde{e}}m_Dm_M)^2}}{(m_{\widetilde{e}}m_Dm_M)}.
\eeq
Three eigenstates of neutral fermions, $(N_1,N_2,N_3)$, are linear combinations of 
$\widetilde{\nu}_L$, $\widetilde{\nu}_L^c$, $n_L$,  defined as 
\beq 
\begin{pmatrix}  \widetilde{\nu}_L \\ \widetilde{\nu}^c_R  \\ n_L \end{pmatrix} = \sum_{a=1}^3 \frac{1}{\sqrt{1+\left ( \frac{m^2_D -m^2_a}{m_D m_{\widetilde{e}}+m_M m_a} \right )^2+\left ( \frac{m_Dm_M + m_{\widetilde{e}} m_a}{m_D m_{\widetilde{e}}+m_M m_a}  \right )^2}}
  \begin{pmatrix} - \frac{m^2_D -m^2_a}{m_D m_{\widetilde{e}}+m_M m_a} \\  \frac{m_Dm_M + m_{\widetilde{e}} m_a}{m_D m_{\widetilde{e}}+m_M m_a} \\ 1  \end{pmatrix}  P_L  N_a
\label{mixingneutral}
\eeq
where $P_L$ is the projection operator, $P_L=(1-\gamma_5)/2$.

Defining $Z^{\rm{ex}}_2 \equiv Z_2^{\psi} \times (-1)^{2 s}$ or $Z_2^{\chi} \times (-1)^{2 s}$ with $s$ 
being the spin of the particle, we can assign the odd $Z^{\rm{ex}}_2$  charge to all the exotic fermions. 
This remnant $Z_2$ symmetry guarantees the stability of the lightest particle among the exotic fermions, 
so that the lightest neutral particle $(\equiv X)$ among $N_a$ could be 
a good cold dark matter candidate.

The extra charged lepton, $\widetilde{e}$, should decay in order to avoid  a stable charged particle.
After the EW symmetry breaking,  the gauge interactions of $\widetilde{e}$ and $N_a$ are described 
in appendix~\ref{gaugeinteraction}.    If $\widetilde{e}$ is heavier than at least $X$,  the charged exotic lepton 
$\widetilde{e}$ decays to the DM $X$ and the SM fermions through the $W^{(*)}$ exchange.

The exotic leptons $\tilde{e}$ and $N_a$ can be produced at colliders by 
DY processes through the $s$-channel $W^\pm, \gamma$ and/or 
$Z^0, Z_H$ exchanges. Note that the DY process through $Z_H$ exchange 
is a new aspect in our model, 
since $Z_H$ couples both to the SM quarks and exotic fermions. 
Once they are produced at colliders, they will decay through the mixing in 
Eq.~(\ref{gauge interaction}) (and the higher-dimensional operators), 
the extra leptons and quarks decay as
\beq
N_{a=2,3} \to X + Z^{0(*)} , \ \ \widetilde{e}^\pm \to N_{a=1,2,3} + W^{\pm (*)} , \ \ 
\eeq
and $W^\pm$ or $Z^0$ (either real or virtual) will decay into two SM fermions. 
Therefore, their collider signatures would be similar to those of charginos and 
neutralinos in supersymmetric models, and bounds on chargino and neutralinos  
could be applied to our model with simple modification. 
The lower bound on $m_{\widetilde{e}}$ would be around $800$ GeV, 
inferred from $pp \to    \chi^{\pm} \chi^0$, $ \chi^{\pm} \chi^{\pm}$ \cite{Bound-ExtraLepton,Bound-ExtraLepton2}. 

\subsection{Extra Quarks}
The SM charges of extra quarks, $q^i_{I}$ $(I=L,R)$, are the same as  those of the right-handed down 
quarks, $D^i_R$.  However they can be distinguished by the $U(1)_H$ charges and $Z^{\rm ex}_2$.
In fact, the mass mixing between the extra quarks and the SM quarks in Eq. (\ref{eq:mass mixing}) 
is forbidden by the symmetry,   
so that the tree-level FCNCs involving the extra quarks are absent and  the exotic quarks can not decay 
at the renormalizable level, in the 2HDM$_{U(1)}$ with $Z^{\rm ex}_2$.   There might be $Z^{\rm ex}_2$ 
symmetric higher-dimensional operators. For example,   a dim-8 operator such as 
\beq \label{higher}
\frac{c_{ijkl}}{\Lambda^4}  \Phi \ov{q_L}^i D_R^j   \ov{l_L}^k E_R^l H_1 + h.c.,
\eeq
would make the extra quarks decay into the SM fermions and the DM $X$, with 
the decay width given by  
\[
\Gamma \sim \frac{1}{(4\pi)^3} \left( \frac{v \cos \beta v_\Phi m_{q}^2}{\Lambda^4} \right)^2 
m_{q},
\]
where $m_q$ is the mass of the exotic quark. 
Assuming $v_\Phi=m_{q}=1$ TeV and $\tan \beta \approx 1$, the lifetime is estimated as
$\sim 1~\mu$sec at $\Lambda=100$ TeV,   which is much longer than the QCD time scale 
$\tau_{\rm QCD} \sim \Lambda_{\rm QCD}^{-1}$.    Therefore they will be  hadronized, forming exotic  
massive hadrons, and would decay inside or outside the detector, depending on its velocity.  
Thus these exotic quarks would be constrained by exotic massive particle searches.
In case they decay inside the detector, the usual bounds from squark search will apply. 

Exotic quarks $q_L^i$ and $q_R^i$ are produced copiously by QCD processes at hadron 
colliders, and by the DY process at lepton colliders. Once they are produced, they will decay
through $Z^{\rm ex}_2$ symmetric higher dimensional operators such as Eq.~(\ref{higher}). 
The extra-quark production could be constrained by the search for squark at LHC 
\cite{Bound-ExtraQuark},  and the lower bound would be also around $1$ TeV.
Once they are produced at colliders, they will decay into the exotic leptons and the SM quarks through 
the mixing in Eq.~(\ref{higher}), and then the  exotic lepton decays into $l_{{\rm SM}} X$ 
through the gauge interaction in Eq.~(\ref{gauge interaction}):
\beq
q_I^i \to q_{{\rm SM}} l_{{\rm SM}} \widetilde{e} ~~~
{\rm followed ~by} ~~~~  \widetilde{e}^\pm \to N_{a=1,2,3} + W^{\pm (*)} ,
\eeq
where $q_{SM}$ and $l_{SM}$ are the SM quarks and leptons.
Therefore the collider signatures will be $4 l + 2 j +  /\!\!\!\!{E_T}$, 
$3 l + 4 j + /\!\!\!\!{E_T}$,  or
$2 l + 6 j + /\!\!\!\!{E_T}$.

\section{Theoretical and Experimental Bounds}
\label{section4}

In this section, we discuss theoretical and experimental constraints  on the 2HDM.

\subsection{Parameters}
In this subsection, we list the parameters in our model which will be scanned over.
In the Higgs potential Eq. (\ref{potential}), there are 11 parameters,
$m_i^2$ ($i=1,2,\Phi$), $\lambda_j$ ($j=1,2,3,4,\Phi$), $\tilde{\lambda}_k$
($k=1,2$), and $\mu$, two of which are fixed by the mass of the SM-like
Higgs boson ($h$), $m_h=125$ GeV, and $v=\sqrt{v_1^2+v_2^2}=246$ GeV.
In the numerical analysis we trace these parameters in the Higgs potential with more 
physical ones related with observables such as masses and the mixing angles: 
\begin{itemize}
\item $\tan\beta = v_2/v_1$, where $v_1$ and $v_2$ are VEVs of $H_i$,
\item $m_A$ : the mass of the pseudoscalar boson,
\item $m_{\tilde{h}}$ : the mass of the  additional neutral Higgs boson due to introducing a new scalar $\Phi$,
\item $m_{H^+}$ : the mass of the charged Higgs boson,
\item $\Delta m_H = m_H - m_A$ : the mass difference between $H$ and $A$,
\item $\alpha,\alpha_1,\alpha_2$ : the mixing angles between three neutral scalar bosons,  
\item $M_{Z_H}$ : the mass of the $U(1)_H$ gauge boson.
\end{itemize} 
The parameters in the Higgs potential can be obtained in terms of these 9 physical parameters. 
For example, $h$, $H$, and $\tilde{h}$ are the physical neutral Higgs bosons,
which are mixtures of three neutral Higgs components of $H_1$, $H_2$, and
$\Phi$, with the mixing angles $\alpha, \alpha_1,\alpha_2$.
The explicit relations are shown in Ref.~\cite{Ko-2HDMtype1}.
The ranges of the parameters are chosen as
$1\le \tan\beta \le 100$, 
$125$ GeV $\le m_A, m_{\tilde{h}} \le 1$ TeV, 
$360$ GeV $\le m_{H^+} \le 1$ TeV, 
$ |\Delta m_H| \le 500$ GeV, 
$|\alpha, \alpha_1, \alpha_2|\le \pi/2$, 
and   $125$ GeV $\le M_{Z_H} \le 1$ TeV. 

$U(1)_H$ gauge interaction is parameterized by the gauge coupling $g_H$ and the $U(1)_H$ gauge 
boson mass $(M_{Z_H})$.   The range of $g_H$ is taken to be $0\le g_H \le 1$ because small $g_H$ is
preferred due to the constraints from EWPOs as we will discuss later.
We note that the VEV of $\Phi$ is obtained by
\begin{equation}
v_\phi = \sqrt{M_{Z_H}^2/g_H^2-v^2 \sin^2\beta}.
\end{equation}
Note that the $Z_H$ mass is bounded from below by 
\[
M_{Z_H} \geq g_H v | \sin\beta | .
\]

In the Yukawa sector for extra chiral fermions, there are four parameters,
$y^q$, $y^l$, $y^n$, and $y^{\prime n}$ as shown in Eq.~(\ref{eq:lepton mass}).
The Yukawa couplings are taken to be in the range,  
\[
0 \le y^q, y^l, y^n, y'^{n} \le 4 \pi .
\]
The masses of extra chiral fermions can be calculated in terms of the Yukawa couplings
($y^q$, $y^l$, $y^n$, $y'^{ n}$) and 3 scalar VEVs ($v_i$ and $v_\phi$).

In summary, we scan 14 parameters in the numerical analysis in order to find 
the regions that are consistent with  theoretical and experimental constraints. 
The mass of the candidate for CDM is not restricted in the discussion
on Higgs physics or electroweak precision tests in Secs.~\ref{section4.3} 
and \ref{section5}.

\subsection{Theoretical Bounds}

In the analysis, we impose perturbativity bounds on dimensionless quartic
couplings in the Higgs potential, $|\lambda_i| \le k$, which are
required for the model to be stable under higher-order corrections.
Here $k$ is a certain number and chosen as $4\pi$ in this work.
The $2\to 2$ scattering processes for scalar and gauge bosons are dominated by the induced quartic 
couplings $Q_i$ at very high energy while the amplitudes including triple gauge couplings are suppressed.
We impose the perturbative unitarity condition on the induced quartic couplings at the tree level with
$|Q_i|\le 8 \pi$~\cite{unitarity}.

Finally, we impose the vacuum stability bounds at the tree level, 
which require that  
the dimensionless couplings $\lambda_{1,2,3,4}$ are to satisfy the following conditions: 
\beq\label{eq:vacuum-condition}
\lambda_1 >0 ,~ \lambda_2>0 ,~ \lambda_3  > - \sqrt{\lambda_1 \lambda_2} ,
~ \lambda_3 +\lambda_4  > - \sqrt{\lambda_1 \lambda_2},
\eeq
in the $\langle \Phi \rangle = 0 $ direction.
They correspond to the ones in the usual 2HDMs without $\lambda_5$.  
It is noticeable that the conditions in Eq.~(\ref{eq:vacuum-condition}) 
lead the scalar mass relation 
\begin{equation}\label{eq:relation-2HDM-0}
m_h^2 +m_H^2-m_A^2 > 0.
\end{equation}
In the ordinary 2HDMs with softly broken $Z_2$ symmetry,
sizable $\lambda_5$ is allowed and 
the conditions (\ref{eq:vacuum-condition}) and (\ref{eq:relation-2HDM-0})
should be modified by the replacements,
$m_A \to m_A +\lambda_5 v^2$ and $\lambda_4 \to \lambda_4+\lambda_5$ 
in Eqs. (\ref{eq:vacuum-condition}), and (\ref{eq:relation-2HDM-0}).

In the $\langle \Phi \rangle \neq 0$ direction, the vacuum-stability conditions for $\lambda_{\Phi}$, $\widetilde{\lambda_1}$ and $\widetilde{\lambda_2}$ are
\begin{eqnarray}
\lambda_{\Phi}>0,~\lambda_1 > \frac{\widetilde{\lambda_1}^2}{\lambda_{\Phi}},~ \lambda_2&>&\frac{\widetilde{\lambda_2}^2}{\lambda_{\Phi}},  ~\lambda_3 -  \frac{\widetilde{\lambda_1}\widetilde{\lambda_2}}{\lambda_{\Phi}} > - \sqrt{ \left( \lambda_1-\frac{\widetilde{\lambda_1}^2}{\lambda_{\Phi}} \right) \left( \lambda_2-\frac{\widetilde{\lambda_2}^2}{\lambda_{\Phi}} \right)}, \nonumber \\
\lambda_3+\lambda_4  -  \frac{\widetilde{\lambda_1}\widetilde{\lambda_2}}{\lambda_{\Phi}}& >& - \sqrt{ \left( \lambda_1-\frac{\widetilde{\lambda_1}^2}{\lambda_{\Phi}} \right) \left( \lambda_2-\frac{\widetilde{\lambda_2}^2}{\lambda_{\Phi}} \right)},
\end{eqnarray}
where the directions of $H_1$ and $H_2$ fields in the last four conditions 
are the same as those of $H_1$ and $H_2$ fields in Eq.~(\ref{eq:vacuum-condition}).

\subsection{Experimental Constraints}
\label{section4.3}

In this subsection, we discuss various experimental constraints  on our 2HDM 
from collider experiments, flavor physics and the electroweak precision observables (EWPOs).

\subsubsection{Electroweak precision observables (EWPOs)}

In order to evaluate the allowed region for the new physics contributions to the EWPOs,
Peskin-Takeuchi parameters $S$, $T$, and $U$ are often used~\cite{peskin}, whose definitions can
be found in Ref.~\cite{EWPO-PDG}. 
According to the recent LHC results, the bounds on $S$, $T$, and $U$
parameters are given by
$S=0.03 \pm 0.10,~T=0.05 \pm 0.12, ~ U=0.03 \pm 0.10,$
with $m^{\rm ref}_h=126$ GeV and $m^{\rm ref}_t=173$ GeV~\cite{Baak:2012kk,Baak:2013ppa}.
The correlation coefficients are $+0.89_{ST}$, $-0.54_{SU}$, and $-0.83_{TU}$.\footnote{Fixing $U=0$, 
$S=0.05 \pm 0.09$ and $T=0.08 \pm 0.07$ with the correlation coefficient $+0.91$.
}
The Peskin-Takeuchi parameters have been calculated in the 2HDM (with extra scalars)~\cite{EWPO-2HDM,EWPO-2HDM2,EWPO-2HDMwScalars1,EWPO-2HDMwScalars2}
and in the 2HDM$_{U(1)}$ (with extra scalars and $U(1)$ gauge boson)~\cite{Ko-2HDMtype1}. 

In addition to the contributions of extra scalar bosons and $U(1)_H$ gauge boson,
there may exist additional contributions from extra fermions in the type-II 2HDM$_{U(1)}$. 
Since the extra quarks are $SU(3) \times U(1)_Y$ vector-like and $SU(2)$ singlet, 
they do not contribute to the EWPOs.
The extra charged leptons and two of three-type neutral leptons are $SU(2) \times U(1)_Y$ vector-like, 
while the other $n_L$ is $SU(2) \times U(1)_Y$ singlet.
The extra neutral leptons mix with each other (see Sec.~\ref{extraleptons}), and they contribute to the EWPOs.
The detail of the extra contribution to the vacuum polarization is shown in Appendix~\ref{ewpos}.

As discussed in Ref. \cite{Ko-2HDMtype1}, $Z_H$ contributes to the EWPOs at tree level through 
the mass mixing between $Z_H$ and $Z$, because the Higgs doublet charged under $U(1)_H$ 
breaks not only the EW symmetry but also the $U(1)_H$ symmetry.
The present authors  discussed the $Z_H$ correction to the EWPOs up to the one-loop level in Ref.~
\cite{Ko-2HDMtype1}.  It is found that the $U(1)_H$ gauge coupling ($g_H$) and  the gauge boson  mass
($M_{Z_H}$) are strictly constrained in the low $Z_H$ mass region especially around the $Z$ boson mass.

In the usual 2HDM, there are two massive CP-even scalars, one massive CP-odd scalar, 
and one charged Higgs pair  after the EW symmetry breaking \cite{Branco:2011iw}. 
They contribute to the EWPOs at the one-loop level,   and it is found that the mass differences among the extra scalars are especially  constrained strongly \cite{EWPO-2HDM,EWPO-2HDM2,Ko-2HDMtype1}. 

In the 2HDM$_{U(1)}$, there is another extra neutral scalar $\tilde{h}$.  
In total, there are three neutral scalar Higgs bosons, $h$, $H$, and $\tilde{h}$ plus one pseudoscalar boson $A$,   where $h$ is the SM-like Higgs boson~\cite{Ko-2HDMtype1} which has 
been observed at the LHC. All scalar bosons contribute to the EWPOs at the one-loop level.
The $U(1)_H$ gauge boson $(Z_H)$ also contributes to the EWPOs,
but its contribution appears even at the tree level through the mixing between
the $\hat{Z}$ and $\hat{Z}_H$ mixing, where $\hat{Z}$ and $\hat{Z}_H$
are gauge eigenstates while $Z$ and $Z_H$ are mass eigenstates, respectively.
In the type-I 2HDM$_{U(1)}$, the contribution of extra scalars and $Z_H$
is discussed in Ref.~\cite{Ko-2HDMtype1}.
In the type-II case, the correction of the scalars and $Z_H$ boson
to the EWPOs is the same as in the type-I case up to the one loop level, 
but there are additional contributions from the extra chiral fermion loops.
They may affect the EWPOs through the self-energy diagrams of the SM gauge bosons.
The formulas for such extra contributions to the EWPOs will be 
given in Appendix~\ref{ewpos} with detailed analysis.

\subsubsection{Constraints on the Charged Higgs boson}
The charged Higgs boson is constrained by direct production channels
in many experiments. In the type-II 2HDM case,
the lower bound for the mass of the charged Higgs boson is about 80 GeV
at the 95\% C.L.~\cite{chargedLEP}.
At the LHC, the stringent bound for $m_{H^+}$ comes from
search for the charged Higgs boson in the top quark decay
for $m_{H^+} < m_t$
and from the direct production of the charged 
Higgs boson with subsequent decays $H^+\to \tau \nu$ or $H^+\to t \bar{b}$
for $m_{H^+} > m_t$~\cite{chargedLHC}.
It is found that the large $\tan \beta$ region is strongly constrained
for $m_{H^+} \lesssim 300$ GeV from the LHC experiments. 

The most stringent bound for $m_{H^+}$ comes from flavour physics,
in particular, $b\to s \gamma$ decays.
In the type-II 2HDM, the region of $m_{H^+}\ge 360$ GeV is allowed
at 95\% C.L.~\cite{Hermann:2012fc}. We adopt this bound in this work.
The $B\to \tau\nu$ decays may constrain $\tan\beta$ and $m_{H^+}$
in the type-II 2HDM. We impose the condition on the branching ratio
for $B\to \tau\nu$ decays, $0.447\times 10^{-4} \le \textrm{Br}(B\to \tau\nu)
\le 1.012 \times 10^{-4}$, which was measured at the Belle with hadronic tagging
for the $\tau$ decay~\cite{btotaunu}. 
The other measurements for the branching ratios for $B\to \tau\nu$ decays 
at the Belle and the BABAR have
much larger uncertainties than the above value~\cite{otherbtotaunu}.
We note that the results in this work do not change so much 
even though we use other results or the average of all results.
The $B_q$-$\bar{B}_q$ mixing is also affected by the charged Higgs exchange.
It is known that the $B_q$-$\bar{B}_q$ mixing disfavors 
a small $\tan \beta$ region so that we impose $\tan\beta \ge 1$~\cite{santos}.
The mass of the pseudoscalar boson and $\tan\beta$ are constrained 
by the production of the pseudoscalar
with the subsequent decays into $A\to \tau^+\tau^-$ or 
$A\to \mu^+\mu^-$~\cite{pseudoLHC}. We take into account this constraint
on $m_A$ and $\tan \beta$ in our analysis.

Another interesting measurement which may strongly affect the constraints
on $\tan \beta$ and $m_{H^+}$ is the branching ratio 
for the semileptonic decay $B\to D^{(\ast)}\tau \nu$.
The BABAR measurement for this branching ratio indicates that the SM as well
as the type-II 2HDM would be excluded with $99.8$\% probability~\cite{btodtaunu}.
This problem would require breaking of the so-called Natural Flavor 
Conservation criteria, which could be realized in the flavor-dependent 
$U(1)$ model~\cite{Ko-B}. However, this breaking cannot be achieved
in the 2HDM$_{U(1)}$ and the anomaly cannot be accommodated with this model.
We ignore the experimental constraint from $Br(B\to D^{(\ast)}\tau\nu)$ 
at the BABAR. 

\subsubsection{Constraints on the neutral (pseudo)scalar bosons}
The search for the SM-like heavy Higgs boson would strongly constrains,
in particular, the heavy Higgs boson mass and its couplings. 
The main channels for the SM-like heavy Higgs boson search are
the $H \to ZZ \to 4 l$ decays in the vector boson fusion (VHF) and
vector boson associated production (VH) or in the $gg$ fusion process ($gg$).
We impose the upper bound on  the signal strength ($\mu$) for a heavy Higgs
boson production and decay:
$\mu_\textrm{VHF+VH}^{ZZ}, \mu_\textrm{gg}^{ZZ} \lesssim 0.1 \sim 1$ 
for $125$ GeV $< m_H < 1$ TeV~\cite{heavyHiggs}. 

The lower bounds on the masses of extra quarks and charged leptons are
set to be 1 TeV and 800 GeV, respectively, as discussed in the previous section.
Finally, there is no bound on the mass of extra neutral leptons, $N_i$,
where the lightest one is a candidate for CDM, $X$.

If $m_X$ is less than $m_h/2$, the observed Higgs boson $h$ can decay to $2X$, which contributes to the invisible decay of $h$.   The bound on the invisible decay of the SM-like 
Higgs has been discussed  in Refs.~\cite{invisible,global}.   
Explicitly we assume $\textrm{BR}(h\to \textrm{invisibles}) \le 0.58$.
We take the mass of the extra scalars and gauge boson to be over the mass of  the SM-like 
Higgs boson.  Thus, they do not contribute to the invisible decay  of $h$.   
Furthermore, if $m_X$ is lighter than the half of the $Z$-boson mass,   $Z$ can also decay 
to $2X$. This constraint may easily be avoided in the range $m_X\ge M_Z/2$.

\begin{center}
\begin{table}
\begin{tabular}{|c|c|c|}
\hline   Higgs tagging channels & ATLAS & CMS \\
\hline   $ H \rightarrow \gamma \gamma$ & $1.57^{+0.33}_{-0.28}$  & $1.13 \pm 0.24$ \\
$H \rightarrow Z Z^*$ & $ 1.44^{+0.40}_{-0.35}$ & $1.00 \pm 0.29$ \\ 
$H \rightarrow W W^*$ & $1.00^{+0.32}_{-0.29}$ & $0.83 \pm 0.21$   \\
$H \rightarrow b\overline{b}$ & $0.2^{+0.7}_{-0.6}$ & $0.93 \pm 0.49$ \\
$H \rightarrow \tau^+ \tau^-$  & $1.09^{+0.36}_{-0.32}$ & $0.91 \pm 0.27$ \\ \hline 
\end{tabular}
\caption{Higgs signal strength data reported at ICHEP2014}
\label{table:signal}%
\end{table}
\end{center}

\begin{figure}[!t]
\begin{center}
{\epsfig{figure=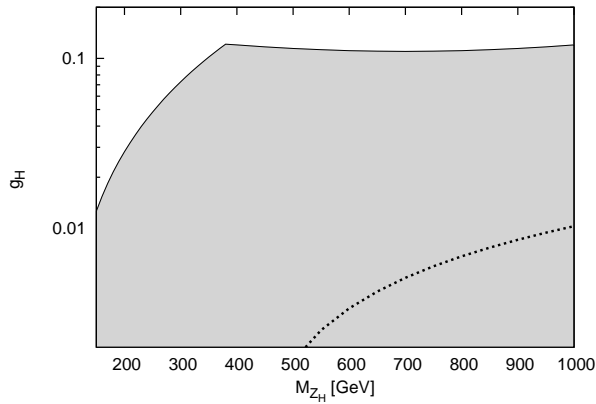,width=0.5\textwidth}}
\end{center}
\vspace{-0.5cm}
\caption{$M_{Z_H}$ and $g_H$ in the type-II 2HDM$_{U(1)}$. The dot line 
is the upper bound on the $U(1)_{\psi}$ gauge boson, 
and the gray region is allowed for the $U(1)_H (\equiv U(1)_b)$ gauge boson.
}
\label{figure1}
\end{figure}

\subsubsection{$U(1)_H$ gauge boson $Z_H$}
On the other hand, the $Z_H$ interaction is constrained by searches for a $Z'$ boson at collider 
experiments.  From now on we define the $U(1)_H$ charge assignments as the leptophobic case, i.e. 
we consider the case $U(1)_H \equiv U(1)_b$. 
Then the $U(1)_b$  gauge interactions of the SM particles are given by   
\begin{eqnarray}
\Hat{{\cal L}}_g &=& g_H \Hat{Z}_H^{\mu} \left (\frac{2}{3} \ov{U_R^i } \gamma_{\mu} U_R^i-\frac{1}{3}\ov{D_R^i } \gamma_{\mu} D_R^i- \frac{1}{3}\ov{Q^i } \gamma_{\mu} Q^i +\ov{N_R^i } \gamma_{\mu} N_R^i \right )+g_Z \Hat{Z}^\mu J^{SM}_\mu \nonumber  \\
&&+\frac{1}{2}M^2_{Z_H}\Hat{Z}_H^{\mu}\Hat{Z}_{H \, \mu} +\frac{1}{2}M^2_{Z}\Hat{Z}^{\mu}\Hat{Z}_{ \mu} +\Delta M^2\Hat{Z}_H^{\mu}\Hat{Z}_{ \mu},
\end{eqnarray}
 where $M_Z^2$, $g_Z$ and $J^{SM}_{\mu}$ are the mass,  the gauge coupling and  the current  of 
 the $Z$ boson in the SM.  The nonzero VEV of $H_2$ gives the mass mixing $\Delta M^2$ between 
 $\Hat{Z}_{H \, \mu}$ and $\Hat{Z}_{ \mu}$, so that $\Hat{Z}_{H \, \mu}$ and $\Hat{Z}_{\mu}$ are not in their 
mass basis.
However, the mixing is strongly constrained by the EWPOs, as discussed in Ref. \cite{Ko-2HDMtype1},
so that $\Hat{Z}_{H \, \mu}$ and $\Hat{Z}_{\mu}$ could be approximately interpreted as the gauge bosons
($Z_{H \, \mu}$ and $Z_{\mu}$)   in their mass basis. 
 Then we can ignore the $Z_H$ boson couplings to  the SM leptons, thereby $Z_H$ becoming leptophobic.
 The strong bounds from the Drell-Yan processes 
and the LEP experiment can be evaded if the mass mixing of $Z$ and $Z_H$ is  small enough.
The resonance searches for a $Z'$ boson in the dijet and $t \ov{t}$ production 
also provide relevant constraints on the $Z_H$ boson. 
They give the upper bound of $g_H$ in the $O(100)$ GeV 
mass region~\cite{dijetbound,dijetbound2,ttbarbound,ttbarbound2}.
In Fig. \ref{figure1}, we depict the allowed region for $g_H$ and $M_{Z_H}$ in the type-II 2HDM$_{U(1)}$ 
with leptophobic $U(1)_H$ ($\equiv U(1)_b$) symmetry,  which is represented by gray color.
For comparison, we also show the upper bound for the $U(1)_{\psi}$ gauge boson,  which is represented 
by the dot line.   The bound for the $U(1)_\psi$ gauge boson is much stronger than that
for the $U(1)_H$ gauge boson due to the interaction with SM leptons.  For the $U(1)_H$ gauge boson, 
it is found that  the low mass region is strictly constrained by the EWPOs, 
i.e. the $Z$ decay width and $\rho$ parameter.   While the bound in high mass region comes mainly
from the resonance searches in the dijet and $t \ov{t}$ production  at hadron colliders.
The allowed value for $g_H$ is $O(0.01)$ in the low $M_{Z_H}$ region  and $O(0.1)$ in the high 
$M_{Z_H}$ region, respectively.   We note that these upper bounds are a bit stronger than 
in the Type-I 2HDM$_{U(1)_H}$ because the $Z_H$ boson is 
fermiophobic in the Type-I case~\cite{Ko-2HDMtype1}.

\section{Higgs signals at the LHC}
\label{section5}

The SM-like Higgs boson with mass $\sim 125$ GeV was discovered at the LHC ~\cite{higgsdiscovery}.
At the first stage of the measurements,  the signal strengths in the Higgs decaying into two photons or 
$ZZ^*$  were slightly larger than the SM predictions.  As more data were accumulated at the 8 TeV 
center-of-momentum (CM) energy,  however,  the signal strengths became consistent with the SM 
predictions in each decay mode as shown in Table~\ref{table:signal}.  
Although this consistency may imply that the discovered boson is really the SM Higgs,  we still cannot  
rule out a possibility it could be  a SM-like Higgs boson in the model  with an extended Higgs sector like 
the 2HDM, with a small mixture from the extended Higgs sector.

In the usual Type-II 2HDM, there are two CP-even scalar bosons,  while there is one more CP-even scalar boson in the 2HDM$_{U(1)_H}$. The lightest one is assumed to be the SM-like Higgs boson in this work. 
In both models, there is a CP-odd scalar $A$.  The Yukawa couplings of the SM-like Higgs boson and extra 
scalar  boson with the SM fermions depend on the vacuum alignment  of  VEVs of two Higgs doublets.
In the 2HDM$_{U(1)_H}$, the $U(1)_H$ gauge boson $Z_H$ and extra chiral fermions also take part 
in interactions. Therefore all the extra particles in 2HDM$_{U(1)_H}$can change  Higgs physics at the 
LHC.  The charged Higgs boson can contribute to $h\to \gamma\gamma$ and  $h\to Z\gamma$ 
at the loop level~\cite{Ko-2HDMtype1}.   Extra charged and/or colored particles contribute
to $h \to \gamma \gamma, Z \gamma$ and the $h \to gg$ at one loop level.   Furthermore, the SM-like 
Higgs boson may decay to the extra particles,  if the sum of masses of final particles are less than $m_h$. 

\begin{figure}[!t]
\begin{center}
{\epsfig{figure=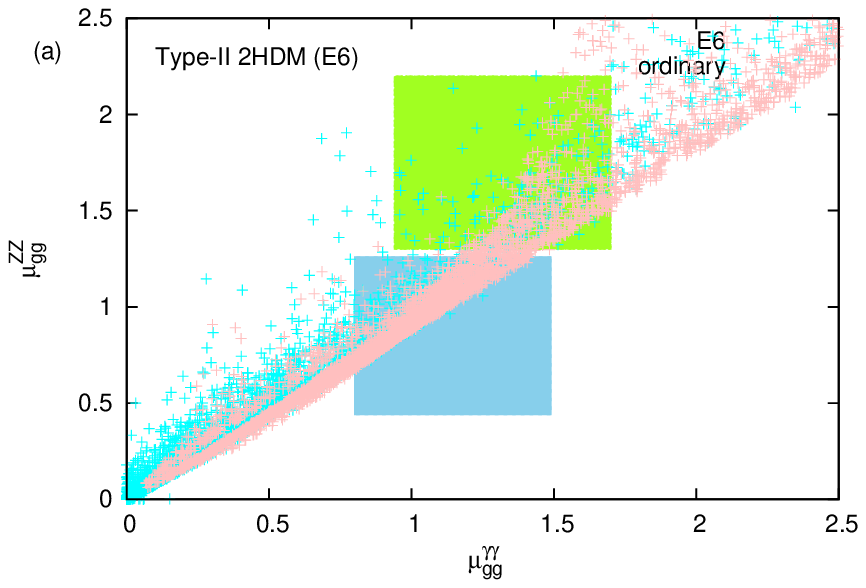,width=0.45\textwidth}}
{\epsfig{figure=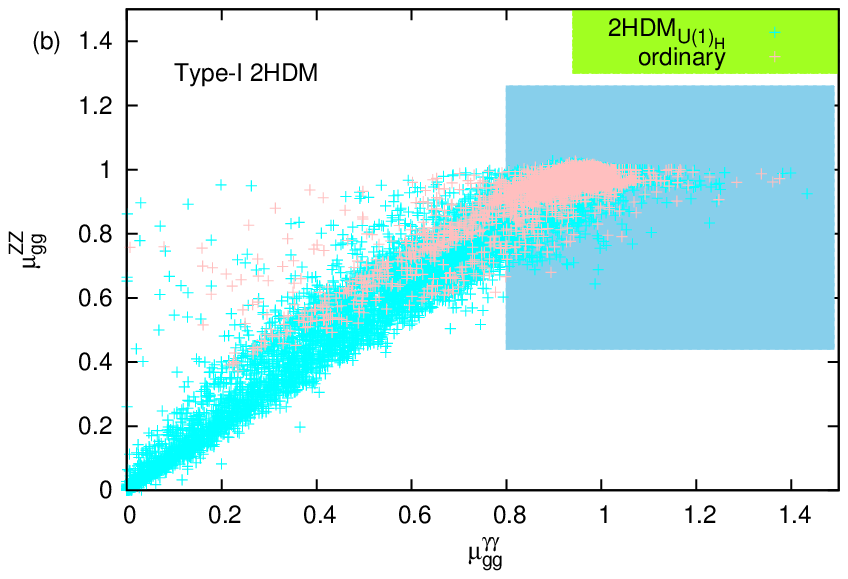,width=0.45\textwidth}}
\end{center}
\vspace{-0.5cm}
\caption{Signal strengths $\mu_{\gamma\gamma}^{gg}$ and $\mu_{ZZ}^{gg}$
(a) in the Type-II 2HDMs and (b) in the Type-I 2HDMs.
}
\label{fig2}%
\end{figure}

In Fig.~\ref{fig2}, we depict the signal strengths $\mu_{\gamma\gamma}^{gg}$ and $\mu_{ZZ}^{gg}$ 
in the Type-II 2HDMs, which are calculated by using HDECAY~\cite{hdecay}.   
We modified the original HDECAY code by modifying Higgs couplings
to the SM fermions and weak gauge bosons and by including the contribution
of the charged Higgs boson and extra charged fermions  to the $h g g$, $h\gamma\gamma$, and 
$h Z\gamma$ vertices.   We also draw the same figure in the Type-I 2HDMs for comparison,
which is based on Ref.~\cite{Ko-2HDMtype1}.    The pink points are allowed in the Type-II (-I) 
2HDM$_{Z_2}$, while the cyan points are allowed in the Type-II (-I) 2HDM$_{U(1)_H}$, respectively.
The blue and green boxes are CMS and ATLAS data at $\sqrt{s}=7, 8$ TeV
in the 1$\sigma$ level, respectively.
Explicitly, we use 
$\mu^{\gamma\gamma}_{ggH}=1.12_{-0.32}^{+0.37}$
and $\mu^{ZZ}_{ggH,t\bar{t}H}=0.80^{+0.46}_{-0.36}$ 
for the CMS data~\cite{cmsnew}
while $\mu^{\gamma\gamma}_{ggF}=1.32\pm 0.38$
and $\mu^{ZZ}_{ggF+b\bar{b}h+t\bar{t}h}=1.7^{+0.5}_{-0.4}$ 
for the ATLAS data~\cite{atlasnew}.

The SM prediction for the Higgs signal strength is $\mu=1$ by definition, which is consistent with 
the CMS and ATLAS  data within $1\sigma$ and $2\sigma$, respectively.
This implies that new physics should not affect the Higgs signal strengths  too much. 
In this respect, the decoupling scenario, where all the scalar bosons except the SM-like 
Higgs boson are heavy enough to decouple from EW physics, or the alignment scenario, where  the heavy 
Higgs boson coupling to gauge boson is suppressed, are preferred in the 2HDMs~\cite{alignment}.
A similar situation would be true in the 2HDM$_{U(1)_H}$. 

\begin{figure}[!t]
\begin{center}
{\epsfig{figure=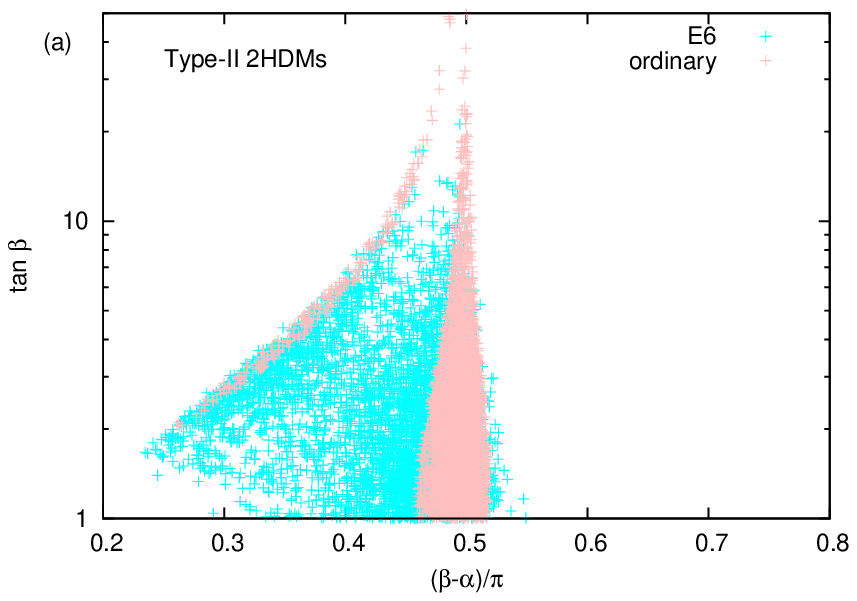,width=0.45\textwidth}}
{\epsfig{figure=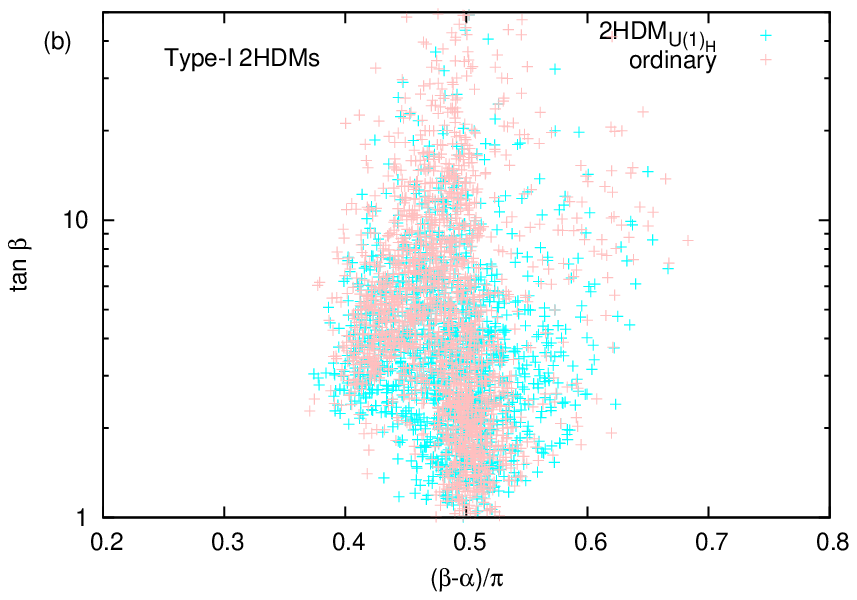,width=0.45\textwidth}}
\end{center}
\vspace{-0.5cm}
\caption{ $(\beta-\alpha)/\pi$ vs. $\tan\beta$
(a) in the Type-II 2HDMs and (b) in the Type-I 2HDMs.
All the points satisfy the CMS data ($\mu_{gg}^{\gamma\gamma}$ and 
$\mu_{gg}^{ZZ}$) within $1\sigma$ level.
}
\label{fig3}%
\end{figure}

In Fig.~\ref{fig3}, we depict the allowed regions in the $( (\beta-\alpha)/\pi, \tan\beta)$ planes for 
(a) the Type-II 2HDMs and   (b) the Type-I 2HDMs,   where all points are consistent with the theoretical 
and experimental bounds discussed in the previous  sections.   We note that the Higgs signal strengths 
$\mu_{gg}^{\gamma\gamma}$ and $\mu_{gg}^{ZZ}$ of all the points in Fig.~\ref{fig3}  are consistent 
with the CMS data ($\mu_{gg}^{\gamma\gamma}$ and $\mu_{gg}^{ZZ}$) in the $1\sigma$ level, as 
shown in Fig.~\ref{fig2}.
If the ATLAS data or combined data of ATLAS and CMS are used,  we would get similar plots.

As shown in Fig.~\ref{fig2}(b), the Higgs signal strengths can reach in the following ranges:
$\mu_{gg}^{\gamma\gamma} \lesssim 1.4$ and
$0.4 \lesssim \mu_{gg}^{ZZ} \lesssim 1.1$ in the Type-I 2HDM$_{Z_2}$,
but $\mu_{gg}^{\gamma\gamma} \lesssim 1.4$ and
$0 \lesssim \mu_{gg}^{ZZ} \lesssim 1.1$ is allowed in the Type-I 2HDM$_{U(1)_H}$.
The region where $\mu_{gg}^{ZZ}\sim 0$ is allowed in the Type-I 2HDM$_{U(1)_H}$,
but it is disallowed in the Type-I 2HDM$_{Z_2}$.
This is because both couplings of the SM-like Higgs boson to fermions and
gauge bosons have an additional suppression factor $\cos\alpha_1$.
That is, the rescaling factors of the SM-like Higgs boson couplings are
$g_{hff}=\cos\alpha_1 \cos\alpha/\sin\beta$ and 
$g_{hVV}=\cos\alpha_1 \sin(\beta-\alpha)$ in the Type-I 2HDM$_{U(1)_H}$.
We note that the rescaling factors in the Type-I 2HDM$_{Z_2}$ can be obtained
if we set  $\alpha_1=0$.

On the other hand, in the Type-II case, both signal strengths 
$\mu_{gg}^{\gamma\gamma}$ and $\mu_{gg}^{ZZ}$ can take their values from 0 to $\sim 3$.
The SM-like Higgs coupling to the SM gauge bosons are the same as in the Type-I  case, but the Yukawa  
couplings are different.  Note that the rescaling factor of the Yukawa coupling to the up-type fermions is
$g_{huu}=\cos\alpha\cos\alpha_1/\sin\beta$, while that to the down-type fermions
is $g_{hdd}=-\sin\alpha\cos\alpha_1/\cos\beta$.

In the Type-I case, the allowed parameter spaces in ordinary 2HDM$_{Z_2}$ and 2HDM$_{U(1)_H}$ 
are rather similar.  As discussed in Ref.~\cite{Ko-2HDMtype1},  $|\sin\alpha|\gtrsim 0.8$ is not allowed 
because the coupling  $g_{hff}\sim \cos\alpha/\sin\beta$ is small for $\tan\beta > 1$.
In this region $|\cos(\beta-\alpha)| \lesssim 0.4$ and the Yukawa couplings  have similar values as 
the SM Yukawa couplings. 

In the Type-II 2HDM$_{Z_2}$, two parameter regions are allowed. One of them is
$(\beta-\alpha)\sim \pi/2$ corresponding to the SM limit line, $\sin(\beta-\alpha)\sim 0$.
The other branch corresponds to the line $\sin(\beta+\alpha) \sim 0$.
In this branch, the Yukawa couplings of the up-type fermions are very close
to the SM Yukawa couplings, while those of the down-type fermions have
the opposite sign relative to the SM Yukawa couplings~\cite{opposite}. 
In the Type-II 2HDM$_{U(1)_H}$, the intermediate region between two pink 
branches is also allowed. This intermediate region contains the parameter space
with $\sin\alpha \sim 0$. The rescaling factor of the Yukawa couplings
of the up-type fermions is $|g_{hu\bar{u}}| \sim 1$, where the opposite sign  is also allowed. 
This is because all the rescaling factors include an overall factor  $\cos\alpha_1$. 
For negative $\cos\alpha_1$, the negative Yukawa coupling  can be achieved.
The rescaling factor of the down-type fermions is allowed  in $|g_{hd\bar{d}}|\lesssim 1$. 
In particular, $|g_{hd\bar{d}}|$ may have  a very small value in some points.
In the analysis, we do not constrain the Yukawa couplings of the down-type  fermions directly. 
If the Yukawa couplings of the down-type fermions are well measured  in the near future, the allowed  
parameter spaces for $g_{hd\bar{d}}$  would strongly be constrained.

In the allowed region, both $\tan\beta$ and $U(1)_H$ coupling are rather small:
$\tan\beta \lesssim 15$ and $g_H \lesssim 0.13$.  There are no strict bounds on the extra scalars, i.e. 
$m_{H,a,\tilde{h}} \ge m_h$ and $ m_{H^+}\ge 360$ GeV.  The mass of the $U(1)_H$ gauge boson is 
in the range of $m_{Z_H}\ge m_h$  and the VEV of $\Phi$ is $v_\phi \gtrsim 2.5$ TeV.   
Because of the small $U(1)_H$ gauge coupling $g_H$, the $Z_H$ boson with $100$ GeV $\sim$ $1$ TeV 
mass  can avoid the strong constraints from experiments.  The mass of the dark matter candidate is 
in the range of $0 < m_X \lesssim 1.2$ TeV.

In principle, our models could be distinguished from the ordinary 2HDMs 
because there exist additional particles: an additional neutral Higgs boson,
a new gauge boson $Z_H$, and extra chiral fermions, that could be produced directly at colliders
or can appear in the loop.   However, note that the qualitative features in Higgs physics, in particular,
the Higgs signal strengths (Fig.~\ref{fig2}) are not so different between two models in the ATLAS/CMS
data regions.   Therefore  it is not that easy to distinguish these two models only by the Higgs signal 
strength measured at the LHC,  since the LHC data are in good agreement with the SM (see Fig.~2).
Large deviations of the Higgs signal strengths from the SM predictions or discovery of new particles 
would be necessary to tell our 2HDM with local $U(1)_H$ gauge symmetry from the usual 2HDM with 
$Z_2$ symmetry.

Still  the detail of the  model parameter space are different as shown in Fig.~\ref{fig3}. 
In the ordinary type-II 2HDM, the allowed region for $\tan\beta$ and $(\beta-\alpha)$ is restricted in the 
two branches,  while in our model, the allowed region is much broader. 
This is mainly due to the  additional neutral Higgs boson with new mixing angles that appear 
in the Yukawa couplings and the Higgs couplings to the weak gauge bosons.   
Also the extra colored and/or charged particles cancelling the gauge anomalies generate the difference 
in the Higgs signal strengths through their contribution to the $h g g$ and $h\gamma\gamma$  couplings.

As we have mentioned, our models have new particles that are not present in the ordinary 2HDM: 
three neutral scalar bosons, a new gauge boson $Z_H$ and extra chiral fermions.  
Discovery of some of these new particles would be distinctive signatures of our model. 
For example, observation of extra fermions in the production/decay channels  discussed in Sec.~III
would be clear signatures of the Type-II 2HDM with local $U(1)_H$ gauge symmetry. 

Note that we have not imposed constraints on the dark matter candidate yet in the analysis of this section. 
The constraints on dark matter from the thermal relic density, direct and indirect detection of dark matter 
will also strongly constrain  the parameter space.   Still  Figs.~\ref{fig2} and \ref{fig3} are meaningful 
if we consider the model where the dark matter candidate can decay by introducing an additional scalar 
that couples with the dark matter candidate. Then there is no dark matter in the model so that we do not 
need to take into account the constraints from dark matter detection.

\section{Dark Matter Physics}
\label{section6}
As we discussed in Sec.~\ref{section4},  the lightest neutral particle $X$ is a Majorana fermion and 
could be stable due to the remnant $Z^{ \rm ex}_2$ symmetry.  In the mass matrix for 
$(\widetilde{\nu}_L,\widetilde{\nu}_R^c,n_L)$,  we assume that $m_{\widetilde{e}} \gg m_D,m_M$ 
in order to evade the stringent constraint from the extra lepton search.  
In that limit, $m_X$ can be approximately evaluated as 
\beq
m_X \approx \frac{2m_Dm_M}{m_{\widetilde{e}}}=\frac{y^n y^{\prime n} v^2}{m_{\widetilde{e}}} \cos \beta \sin \beta ,
\eeq
where $X$ is mostly $n_L$-like: 
\[
X_L \approx n_L -\frac{m_D}{m_{\widetilde{e} }} \widetilde{\nu}_L-\frac{m_M}{m_{\widetilde{e} } }
\widetilde{\nu}_R^c \  .
\] 
In order to make $m_X$ heavy enough, we require large Yukawa couplings, $y^n$ and $y^{\prime n}$ 
of $\sim O(1)$.

\begin{figure}[!t]
\begin{center}
{\epsfig{figure=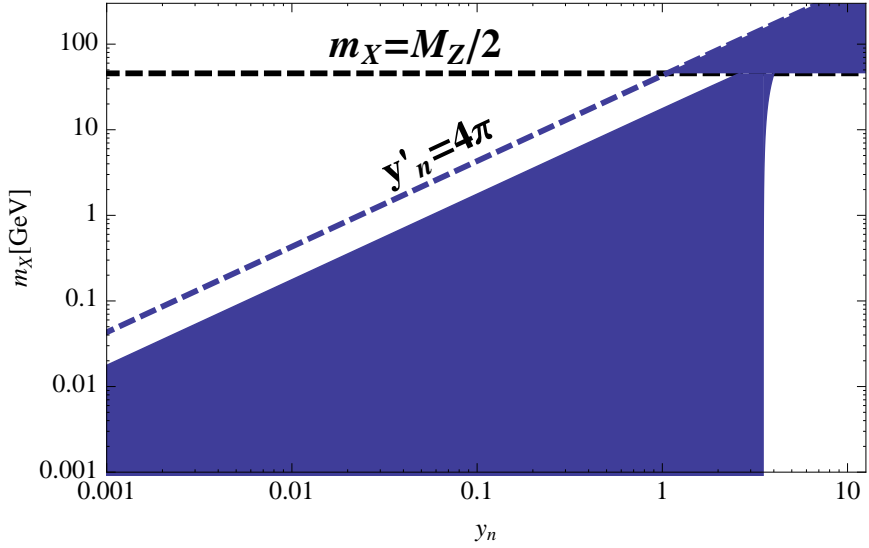,width=0.7\textwidth}} 
\end{center}
\vspace{-0.5cm}
\caption{$y^n$ vs. $m_{X}$ with $\tan \beta=3$ and 
$m_{\widetilde{e}}=1$ TeV.  
The blue region satisfies all the bounds on the extra particles in the text, as well as the invisible $Z$ 
decay  ($y^n \lesssim 3$ and $y^{\prime n} \lesssim 1$). 
}
\label{fig4}
\end{figure}
\begin{figure}[!t]
\begin{center}
{\epsfig{figure=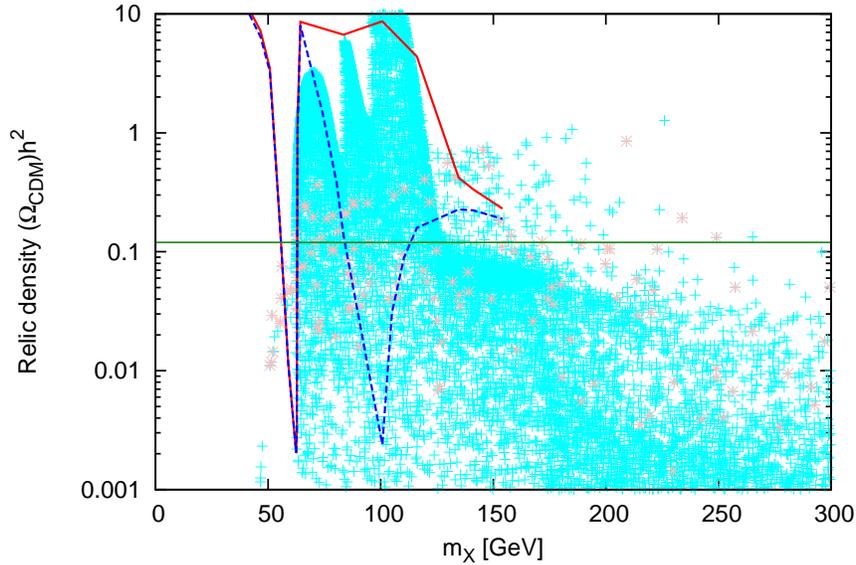,width=0.7\textwidth}}
\end{center}
\vspace{-0.5cm}
\caption{$m_{X}$ vs. $\Omega h^2$ in the decoupling limit,
$\sin(\beta-\alpha)=0, \alpha_1=\alpha_2=0$.
The cyan points satisfy the experimental constraints at colliders
while the pink points satisfy the bound from direct detection of DM
in the LUX experiment too.
The red line is for 500 GeV $\le m_{H,A,H^+} \le$ 1 TeV and $y^{\prime n}=1$,
while the blue line is for $m_A=m_H=200$ GeV, $m_{H^{+}}=360$ GeV and 
$y^{\prime n}=1$.
}
\label{fig5}
\end{figure}

In Fig.~\ref{fig4}, the region for $m_X$ and $y^n$ is described, setting $\tan \beta=3$ 
and $m_{\widetilde{e}}=1$TeV.   The blue region is the one allowed for the bound from the invisible decay 
of $Z$ boson ($y^n \lesssim 3$ and $y^{\prime n} \lesssim 1$), which is derived from the $1 \sigma $ error of 
the invisible decay width of $Z$ boson \cite{PDG}.

$X$ could be thermally  produced through the following annihilation processes: $XX \to f \overline f$, 
$W^+W^-$, and $ZZ$. The extra fermions  masses are generated by $\langle \Phi \rangle$,  and could be 
much heavier  than $X$ because of the experimental constraints, so that they have already decoupled 
at the freeze-out temperature of $X$.  The $U(1)_H$ gauge  interaction  through $Z_H$ emission may 
be also effective as we see in Fig.~\ref{figure1}.  And $h$, $A$, $H$ exchanging in the $s$-channel are 
efficient in the annihilation and scattering with nuclei,   because of large Yukawa coupling. 
However, all the cross sections are strongly suppressed by the mixing elements $U_{ab}$'s 
defined in Eq. (\ref{mixingneutral}) and Appendix \ref{gaugeinteraction}, 
so that $X$ tends to be over-produced in our universe.

In Fig.~\ref{fig5}, we show the thermal relic density of $X$ 
in the decoupling limit where the mixing among 
CP-even scalars are fixed at $\sin(\beta-\alpha)=1$ and $\alpha_1=\alpha_2=0$. 
The cyan points satisfy the experimental constraints at colliders
which have been discussed
in previous sections. The pink points satisfy the LUX bound
for direct detection of DM in addition to the experimental constraints
at colliders. 

For more concrete discussion, let us fix other parameters too.
In the red line, the masses of $A$, $H$ and $H_{\pm}$
are within $500$ GeV $ \leq m_{H,A,H^+} \leq 1$ TeV and $y^{\prime n}=1$.
Then the heavy scalar exchange processes are inefficient to reduce
the relic density for $m_X \lesssim 200$ GeV.
The blue line corresponds to the case with $m_A=m_H=200$ GeV,
$m_{H \pm}=360$ GeV, and $y^{\prime n}=1$.
The green band is the observed relic density in the PLANCK experiment~\cite{planck}.
In both cases, only the regions around the resonances, $m_X \approx m_h/2$ and $m_X \approx m_H/2$, 
can result in the correct relic density of DM (we have calculated thermal relic density using the micrOMEGAs \cite{micromegas}).
The spin-independent and spin-dependent direct detection cross sections of DM X on proton 
are estimated as $\sigma_{\rm SI}=6.54 \times 10^{-10}$ $(1.98 \times 10^{-10})$ pb and 
$\sigma_{\rm SD}=2.41 \times 10^{-8}$ $(1.91 \times 10^{-5})$ pb at $m_X= 55.3$ $(83.5)$ GeV, 
where the DM density is $\Omega h^2=0.166$ $(0.137)$. 
They are far below the current experimental bounds from the direct detection \cite{LUX}.
If $m_X$ is less than half of $m_h$, the SM-like Higgs can decay into a pair of DMs, and
the branching ratio of the invisible decay is $0.1$ at $m_X=55.3$ GeV, 
which is still acceptable~\cite{global}. 

In Fig.~\ref{fig5}, there is a sharp peak around $m_X=50$ GeV,  where the relic density is highly
suppressed due to the processes, $XX\to Z \to f\bar{f}$ ($f=$ SM fermions except $t$), 
(the SM $Z$ boson resonance). The DM coupling with the SM $Z$ boson is  generated by the 
$Z$-$Z_H$ mixing. At $m_X \approx 60$ GeV,  the relic density can be smaller than the current 
observation due to the resonance effect of the SM-like Higgs boson mediation. In the region 
$m_X > 60 $ GeV, new resonance processes, for example, the heavy scalar ($H$) exchange process, 
could contribute  to decreasing the relic density. At $m_X \simeq 80$ GeV,  $XX\to W^+ W^-$ channel is 
open so that the relic density could be below the observation. 
Surely, it strongly depends on the DM-$Z$ coupling.  
For a small DM-$Z$ coupling, the $XX\to W^+ W^-$ process is not sufficient
to reduce the relic density. In that case, the relic density is higher 
than the current observation as shown in Fig.~\ref{fig5}.

\begin{figure}[!t]
\begin{center}
{\epsfig{figure=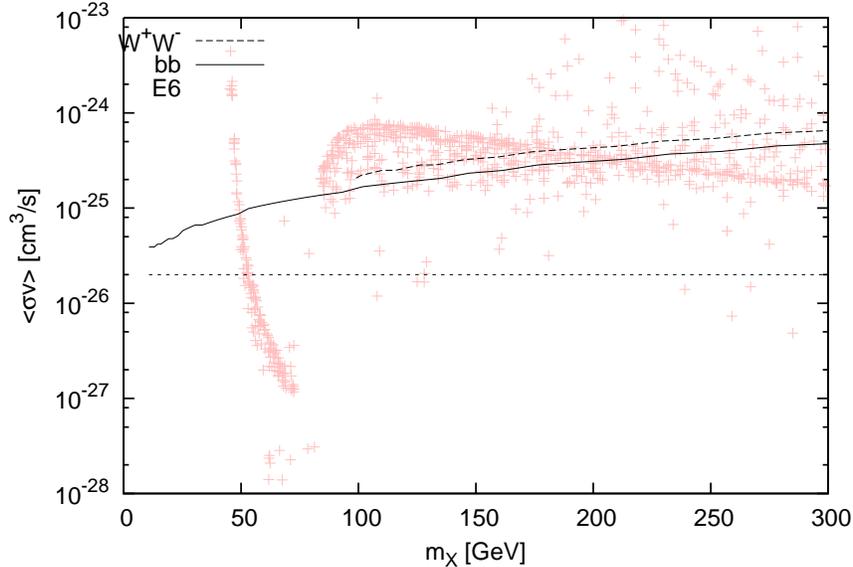,width=0.7\textwidth}}
\end{center}
\vspace{-0.5cm}
\caption{$m_X$ vs. $\langle \sigma v \rangle$ in units of GeV and cm$^3/$s,
respectively.
The pink points satisfy collider constraints and direct detection bound
in the LUX experiments. The relic density is below the current observation
by the PLANCK Collaboration~\cite{planck}. 
}
\label{fig6}
\end{figure}

In Fig.~\ref{fig6}, we depict the DM mass vs. the velocity-averaged annihilation
cross section, $\langle \sigma v \rangle$,
at the halo. The calculation was carried out by using micrOMEGAs~\cite{micromegas}.
All  points satisfy the collider constraints discussed in previous sections
as well as the LUX bound for direct detection of DM.
We impose that the thermal relic densities are below the PLANCK observation
$\Omega_\textrm{CDM}h^2=0.1199\pm0.0027$ with 3$\sigma$ uncertainty~\cite{planck}.
The horizontal line, whose value is $\langle \sigma v \rangle \simeq
3\times 10^{-26}$ cm$^3/$s, corresponds to the bound from the relic density
for the $s$-wave annihilation dominant case. 
The solid and dotted curves are the Fermi-LAT bound for the DM annihilation
into $b\bar{b}$ and $W^+ W^-$, respectively.

At the $Z$-resonance region ($m_X \sim 50$ GeV), the indirect detection bound
might severely constrain our model. The bound is assumed that $XX\to b\bar{b}$
is dominant, but in our model its contribution is about 17\%. 
The rest contribution comes from the DM annihilation into SM fermions
except $b$ and $t$ pairs. This would slightly relieve the strong bound from 
the indirect detection of DM.
As shown in Fig.~\ref{fig6}, the Higgs-resonance region is less constrained
because of the resonance effects coming mainly from difference between
the DM velocity at the freeze-out and at the current halo.
At the region $m_X\gtrsim m_W$, our model is strongly constrained by 
the indirect detection of DM again, but there are still some allowed regions
as shown in Fig.~\ref{fig6}.

\section{Summary}
\label{section6}

In this paper, we have studied the type-II 2HDM$_{U(1)}$ inspired by $E_6$, which was proposed 
by the present authors a few years ago~\cite{Ko-2HDM}. 
Both of the two Higgs doublets ($H_1$ and $H_2$) and the SM chiral fermions are charged under  
$U(1)_H$ and the theory becomes anomalous.  Therefore, for the purpose of anomaly cancellation, 
we have introduced extra quarks and leptons, which  could be derived from ${\bf 27}$ representation 
together with the SM fermions, from the point of view of bottom-up approach.  Unlike the fermion sector,  
the Higgs sector of our model has not been extended to ${\bf 27}$ representation of $E_6$. 
This is the main difference between our model and the usual $E_6$ GUT. 
Still we could expect that our 2HDM$_{U(1)}$ is effectively realized by supersymmetric  $E_6$ GUT. 
But we have simply discussed the phenomenology involving the extra Higgs doublet and fermions, 
assuming only one gauged $U(1)_H$ and $Z^{\rm ex}_2$ discrete symmetry
which may be predicted by $E_6$ gauge symmetry survive at low energy.
Especially, we considered the leptophobic $U(1)_H$ charge assignments  to avoid  the stringent 
constraint from DY processes, and study the $Z_H$ effect on the Higgs and DM physics.   
In fact, $Z_H$ could be as light as $\sim O(100)$ GeV, but very light $Z_H$ is strictly constrained 
by the EWPOs,  as we see in Fig. \ref{figure1}.
It may be difficult to draw explicit bounds on 2HDMs from the Higgs signal strengths at the LHC alone. 
But we found that the Type-II 2HDM can easily enhance or reduce the signal strength of 
$h \to ZZ$ and $\gamma \gamma$, because of the sensitivity to $h \to b \overline b$,  compared with 
Type-I 2HDMs, so that we can expect that our Type-II 2HDM$_{U(1)}$ will be strictly constrained 
by the LHC Run-II.

Search for the extra quarks and leptons is especially important to our model.
The current lower bound from the exotic fermion searches is around $800$ GeV, and
it will become more stringent at the LHC Run-II. We may find Majorana fermion dark matter candidate 
among the extra neutral particles, where the DM stability is guaranteed by $Z^{\rm ex}_2$. 
It may be difficult to shift the mass of the DM, because of the stringent constraints on the extra charged 
particles and $Z^{\rm ex}_2$; $O(10)$ GeV DM mass corresponds to $O(1)$ Yukawa couplings, $y^n$ 
and $y'^n$. If we accept such large Yukawa couplings, we can explain the correct thermal relic DM density,
and escape from the strong bounds from the DM direct detections.  
The DM scenario predicts the invisible decay of the $125$ GeV Higgs and the branching ratio is
$\sim O(0.1)$, which may be reached at the LHC. 
$Z^{\rm ex}_2$ plays two important roles: it does not only guarantee the DM stability, but also forbid the 
unwanted FCNCs involving  the extra fermions. 
In other words, we have to consider the effect of the mixing terms between the SM particles
and the extra fermions as in Eq. (\ref{eq:mass mixing}), if we cannot realize $Z^{\rm ex}_2$
from $E_6$ gauge symmetry. It will cause problems in flavor physics, but
may give some rich phenomenology in neutrino and dark matter physics. 
It was discussed in Ref. \cite{Ko:2014tca} and the detail of the work is in progress.

\acknowledgments

We thank Korea Institute for Advanced Study for providing computing resources 
(KIAS  Center for Advanced Computation Abacus System) for this work.
This work was supported in part by Basic Science Research Program through the
National Research Foundation of Korea (NRF) funded by the Ministry of Education Science
and Technology 2011-0022996 (CY), by NRF Research Grant 2012R1A2A1A01006053 
(PK and CY), and by SRC program of NRF funded by MEST (20120001176)
through Korea Neutrino Research Center at Seoul National University (PK).  
The work of YO is supported by Grant-in-Aid for Scientific research from the Ministry of Education, Science, Sports, and Culture (MEXT), Japan, No. 23104011.



\appendix

\section{Interactions of exotic fermions}
\label{gaugeinteraction}
The gauge interactions of the extra fermions are given by the following terms,
\begin{eqnarray}
{\cal L}_g &=&g_s G^{a\mu} \ov{q} t^a \gamma_{\mu} q- \frac{g'}{3} B^{\mu} \ov{q} \gamma_{\mu} q + g_H \Hat{Z}_H^{\mu} ( Q_{q_L} \ov{q_L} \gamma_{\mu} q_L+ Q_{q_R} \ov{q_R} \gamma_{\mu} q_R) \nonumber  \\
&&+ \frac{e}{4c_W s_W}  \Hat{Z}^{\mu}(|U_{\nu_Ra}|^2-|U_{\nu_La}|^2) (\ov{N}_a \gamma_{\mu} \gamma_5 N_a) \nonumber  \\
&&+\sum_{( a , b)=(a,a+1)}\frac{e}{2c_W s_W}  \Hat{Z}^{\mu} \{ Re(U^*_{\nu_Ra}U_{\nu_Rb})-Re(U^*_{\nu_La}U_{\nu_Lb}) \} (\ov{N}_a \gamma_{\mu} \gamma_5 N_b)  \nonumber  \\
&&+\sum_{( a , b)=(a,a+1)}\frac{ie}{2c_W s_W}  \Hat{Z}^{\mu} \{ Im(U^*_{\nu_La}U_{\nu_Lb})-Im(U^*_{\nu_Ra}U_{\nu_Rb}) \} (\ov{N}_a \gamma_{\mu}  N_b)  \nonumber  \\
&&-\frac{e(c_W^2-s_W^2)}{2c_Ws_W} \Hat{Z}^{\mu} \ov{\widetilde{e}} \gamma_{\mu} \widetilde{e}-eA^{\mu} \ov{\widetilde{e}} \gamma_{\mu} \widetilde{e} \nonumber \\
&&+\frac{e}{s_W 2 \sqrt{2}}W^{\mu \dagger} \{ (U_{\nu_La}+U^*_{\nu_Ra})\ov{\widetilde{e}} \gamma_{\mu} N_a- (U_{\nu_La}-U^*_{\nu_Ra})\ov{\widetilde{e}} \gamma_{\mu} \gamma_5 N_a \} \nonumber \\
&&+\frac{e}{s_W 2 \sqrt{2}}W^{\mu}  \{ (U^*_{\nu_La}+U_{\nu_Ra})\ov{N_a} \gamma_{\mu} \widetilde{e}- (U^*_{\nu_La}-U_{\nu_Ra})\ov{N_a} \gamma_{\mu} \gamma_5 \widetilde{e} \}  \nonumber  \\
&&+\frac{g_H}{2} \Hat{Z}_H^{\mu}(Q_{\widetilde{e}_R}|U_{\nu_Ra}|^2-Q_{\widetilde{e}_L}|U_{\nu_La}|^2-Q_{n_L}|U_{n_La}|^2) (\ov{N}_a \gamma_{\mu} \gamma_5 N_a)  \nonumber  \\
&&+g_H \Hat{Z}_H^{\mu}\sum_{( a , b)=(a,a+1)}  \{ Q_{\widetilde{e}_R} Re(U^*_{\nu_Ra}U_{\nu_Rb})- Q_{\widetilde{e}_L}Re(U^*_{\nu_La}U_{\nu_Lb}) - Q_{n_L}Re(U^*_{n_La}U_{n_Lb})  \} (\ov{N}_a \gamma_{\mu} \gamma_5 N_b)  \nonumber  \\
&&-ig_H \Hat{Z}_H^{\mu}\sum_{( a , b)=(a,a+1)}  \{ Q_{\widetilde{e}_R} Im(U^*_{\nu_Ra}U_{\nu_Rb})- Q_{\widetilde{e}_L}Im(U^*_{\nu_La}U_{\nu_Lb}) - Q_{n_L}Im(U^*_{n_La}U_{n_Lb})  \} (\ov{N}_a \gamma_{\mu}  N_b), \nonumber
\label{gauge interaction}
\end{eqnarray} 
where $\{(a,b) \}=\{ (1,2),(2,3),(3,1) \}$ and $(3,4)=(3,1)$ are defined. 
The flavor index is omitted assuming the masses are degenerate and
the mixing, $U_{ab}$, is defined in Eq. (\ref{mixingneutral}). 

The Yukawa couplings involving neutral fermions may be relevant to dark matter physics:
\begin{eqnarray}
{\cal L}_Y& =&-\frac{m_{\widetilde{e}}}{v_\Phi} (U_{\nu_Ra} U_{\nu_Lb}) \Phi \ov{N^a} P_L N^b -\frac{m_D}{v\cos \beta} (U_{\nu_Ra} U_{n_Lb}) H^0_1 \ov{N^a} P_L N^b  -\frac{m_M}{v\sin \beta} (U_{\nu_La} U_{n_Lb}) H^0_2 \ov{N^a} P_L N^b   \nonumber \\
&&-\frac{m_D \sqrt{2}}{v \cos \beta} U_{n_Lb} H^-_1 \ov{\widetilde{e}} P_L N^b-\frac{m_M\sqrt{2}}{v \sin \beta}  U_{n_Lb} H^+_2  \ov{\widetilde{e}^c} P_L N^b + h.c..
\end{eqnarray} 
$H^0_{1,2}$ and $H^{\pm}_{1,2}$ are the neutral and charged components of the two Higgs doublets
and they generally mix with each other as discussed in Ref. \cite{Ko-2HDMtype1}.

\section{Contribution of the extra lepton to the Vacuum Polarization}
\label{ewpos}
Simply assuming that the masses are degenerate among the generations,
the corrections of the vacuum polarizations are given by
\begin{eqnarray}
\Delta \Pi_{WW}(q^2)&=& \frac{e^2N_F}{8 s^2_W}  \sum_a   \{ |U_{\nu_La}+U^*_{\nu_Ra}|^2 \Pi_V(q^2,m_{\widetilde{e}},m_a) +|U_{\nu_La}-U^*_{\nu_Ra}|^2  \Pi_A(q^2,m_{\widetilde{e}},m_a) \}, \nonumber \\
&& \\
\Delta \Pi_{ZZ}(q^2)&=&\frac{N_Fe^2(c_W^2-s_W^2)^2}{4c^2_W s^2_W}\Pi_{V}(q^2,m_{\widetilde{e}},m_{\widetilde{e}}) \nonumber  \\
&&+ \sum_a  \frac{N_F e^2}{8c^2_W s^2_W} (|U_{\nu_Ra}|^2-|U_{\nu_La}|^2) ^2
\Pi_{A}(q^2,m_a,m_a)   \nonumber  \\
&&+  \sum_{(a,b)=(a,a+1)}\frac{N_Fe^2}{4c^2_W s^2_W}  \{ Re(U^*_{\nu_Ra}U_{\nu_Rb})-Re(U^*_{\nu_La}U_{\nu_Lb}) \}^2
\Pi_{A}(q^2,m_a,m_b)   \nonumber  \\
&&+  \sum_{(a,b)=(a,a+1)}\frac{N_Fe^2}{4c^2_W s^2_W}  \{ Im(U^*_{\nu_Ra}U_{\nu_Rb})-Im(U^*_{\nu_La}U_{\nu_Lb}) \}^2
\Pi_{V}(q^2,m_a,m_b),    \\
\Delta \Pi_{Z \gamma}(q^2)&=&N_F \frac{e^2(c_W^2-s_W^2)}{2c_W s_W}   \Pi_{V}(q^2,m_{\widetilde{e}},m_{\widetilde{e}}), \\
\Delta \Pi_{\gamma \gamma}(q^2)&=&N_F e^2  \Pi_{V}(q^2,m_{\widetilde{e}},m_{\widetilde{e}}),
\end{eqnarray}
where $\Pi^V$ ($\Pi^A$) is the vacuum polarization of vector currents (axial currents) and defined as
\begin{eqnarray}
\Pi^V(q^2,m_1,m_2)&=&\frac{-1}{8 \pi^2} \left \{ (q^2-(m_1-m_2)^2) B_0 (q^2,m_1^2,m_2^2)+4 B_{22} (q^2,m_1^2,m_2^2)-A(m_1^2)-A(m_2^2) \right \},  \nonumber \\
\Pi^A(q^2,m_1,m_2)&=&\frac{-1}{8 \pi^2} \left \{ (q^2-(m_1+m_2)^2) B_0 (q^2,m_1^2,m_2^2)+4 B_{22} (q^2,m_1^2,m_2^2)-A(m_1^2)-A(m_2^2) \right \}.  \nonumber \\
\end{eqnarray}
The explicit expressions of the functions $B_0$, $B_{22}$ and $A$ and
the contribution of the Majorana particles can be found in 
Ref.~\cite{EWPO-2HDM,Kniehl:1992ez}.


\vspace{-1ex}

\begin{thebibliography}{99}




\bibitem{Branco:2011iw} 
  G.~C.~Branco, P.~M.~Ferreira, L.~Lavoura, M.~N.~Rebelo, M.~Sher and J.~P.~Silva,
  Phys.\ Rept.\  {\bf 516}, 1 (2012)
  [arXiv:1106.0034 [hep-ph]].


\bibitem{2HDMLHC}
  G.~Belanger, B.~Dumont, U.~Ellwanger, J.~F.~Gunion and S.~Kraml,
  Phys.\ Rev.\ D {\bf 88}, 075008 (2013)
  [arXiv:1306.2941 [hep-ph]];
  C.~W.~Chiang and K.~Yagyu,
  JHEP {\bf 1307}, 160 (2013)
  [arXiv:1303.0168 [hep-ph]];
  B.~Grinstein and P.~Uttayarat,
  JHEP {\bf 1306}, 094 (2013)
  [Erratum-ibid.\  {\bf 1309}, 110 (2013)]
  [arXiv:1304.0028 [hep-ph]];
  A.~Celis, V.~Ilisie and A.~Pich,
  JHEP {\bf 1307}, 053 (2013)
  [arXiv:1302.4022 [hep-ph]];
  H.~S.~Cheon and S.~K.~Kang,
  JHEP {\bf 1309}, 085 (2013)
  [arXiv:1207.1083 [hep-ph]];
  S.~Chang, S.~K.~Kang, J.~P.~Lee, K.~Y.~Lee, S.~C.~Park and J.~Song,
  JHEP {\bf 1305}, 075 (2013)
  [arXiv:1210.3439 [hep-ph]];
  S.~Chang, S.~K.~Kang, J.~P.~Lee, K.~Y.~Lee, S.~C.~Park and J.~Song,
  JHEP {\bf 1409}, 101 (2014)
  [arXiv:1310.3374 [hep-ph]];
  A.~Celis, V.~Ilisie and A.~Pich,
  JHEP {\bf 1312}, 095 (2013)
  [arXiv:1310.7941 [hep-ph]];
\bibitem{Craig:2013hca}
  N.~Craig, J.~Galloway and S.~Thomas,
  arXiv:1305.2424 [hep-ph].
  P.~M.~Ferreira, R.~Guedes, M.~O.~P.~Sampaio and R.~Santos,
  JHEP {\bf 1412}, 067 (2014)
  [arXiv:1409.6723 [hep-ph]];
  J.~Song and Y.~W.~Yoon,
  arXiv:1412.5610 [hep-ph].


\bibitem{Glashow}
  S.~L.~Glashow and S.~Weinberg,
  Phys.\ Rev.\ D {\bf 15} (1977) 1958.  


\bibitem{Ko-2HDMtype1} 
  P.~Ko, Y.~Omura and C.~Yu,
  JHEP {\bf 1401}, 016 (2014)
  [arXiv:1309.7156 [hep-ph]].

  
\bibitem{Ko-2HDM} 
  P.~Ko, Y.~Omura and C.~Yu,
  Phys.\ Lett.\ B {\bf 717}, 202 (2012)
  [arXiv:1204.4588 [hep-ph]].




 \bibitem{Ko-Top}
  P.~Ko, Y.~Omura and C.~Yu,
  Phys.\ Rev.\ D {\bf 85}, 115010 (2012)
  [arXiv:1108.0350 [hep-ph]].
\bibitem{Ko-Top2}
  P.~Ko, Y.~Omura and C.~Yu,
  JHEP {\bf 1201} (2012) 147 [arXiv:1108.4005 [hep-ph]].




\bibitem{Ko-IDM} 
  P.~Ko, Y.~Omura and C.~Yu,
  JHEP {\bf 1411}, 054 (2014)
  [arXiv:1405.2138 [hep-ph]].

\bibitem{Baek:2013qwa}
  S.~Baek, P.~Ko and W.~I.~Park,
  JHEP {\bf 1307}, 013 (2013)
  [arXiv:1303.4280 [hep-ph]].
  
\bibitem{Baek:2013dwa} 
 S.~Baek, P.~Ko and W.~I.~Park,
  JCAP {\bf 1410}, 067 (2014)
  [arXiv:1311.1035 [hep-ph]].
  
\bibitem{Ko:2014nha} 
P.~Ko and Y.~Tang,
  JCAP {\bf 1405}, 047 (2014)
  [arXiv:1402.6449 [hep-ph]].
  
\bibitem{Ko:2014bka}
P.~Ko and Y.~Tang,
  Phys.\ Lett.\ B {\bf 739}, 62 (2014)
  [arXiv:1404.0236 [hep-ph]].
  
\bibitem{Baek:2014kna} 
S.~Baek, P.~Ko and W.~I.~Park,
  arXiv:1407.6588 [hep-ph].

\bibitem{Ko:2014lsa} 
  P.~Ko and Y.~Tang,
  Phys.\ Lett.\ B {\bf 741}, 284 (2015)
  [arXiv:1410.7657 [hep-ph]].


\bibitem{E6Zprime1} 
  S.~M.~Barr,
  Phys.\ Rev.\ Lett.\  {\bf 55}, 2778 (1985).
  
\bibitem{E6Zprime}
  D.~London and J.~L.~Rosner,
  Phys.\ Rev.\  D {\bf 34}, 1530 (1986).


  
\bibitem{E6review} 
  J.~L.~Hewett and T.~G.~Rizzo,
  Phys.\ Rept.\  {\bf 183}, 193 (1989).
  
\bibitem{E6review2} 
  S.~F.~King, S.~Moretti and R.~Nevzorov,
  Phys.\ Rev.\ D {\bf 73}, 035009 (2006)
  [hep-ph/0510419].

\bibitem{Babu}
  K.~S.~Babu, C.~F.~Kolda and J.~March-Russell,
  Phys.\ Rev.\  D {\bf 54}, 4635 (1996)
  [arXiv:hep-ph/9603212].
\bibitem{Rizzo} 
  T.~G.~Rizzo,
  Phys.\ Rev.\ D {\bf 85}, 055010 (2012)
  [arXiv:1201.2898 [hep-ph]].

\bibitem{E6bound}
CMS Collaboration, CMS-PAS-EXO-12-061.

\bibitem{E6leptophobic}
  J.~L.~Rosner,
  Phys.\ Lett.\  B {\bf 387}, 113 (1996)
  [arXiv:hep-ph/9607207].

\bibitem{E6leptophobic2} 
  K.~Leroux and D.~London,
  Phys.\ Lett.\ B {\bf 526}, 97 (2002)
  [hep-ph/0111246].

\bibitem{E6leptophobic3} 
  M.~R.~Buckley, D.~Hooper and J.~L.~Rosner,
  Phys.\ Lett.\ B {\bf 703}, 343 (2011)
  [arXiv:1106.3583 [hep-ph]].

\bibitem{E6leptophobic4} 
  C.~W.~Chiang, T.~Nomura and K.~Yagyu,
  JHEP {\bf 1405} (2014) 106
  [arXiv:1402.5579 [hep-ph]].



\bibitem{Bound-ExtraLepton}
CMS Collaboration, CMS-PAS-SUS-13-006, CERN, Geneva Switzerland (2013).
\bibitem{Bound-ExtraLepton2}
ATLAS Collaboration,ATLAS-CONF-2013-049, CERN, Geneva Switzerland (2013).
\bibitem{Bound-ExtraQuark}
CMS Collaboration, ATLAS-CONF-2013-089,``First Large Hadron Collider Physics Conference," Barcelona, Spain, 13 - 18 May 2013.




\bibitem{unitarity}
  S.~Kanemura, T.~Kasai and Y.~Okada,
  Phys.\ Lett.\ B {\bf 471}, 182 (1999)
  [hep-ph/9903289];
  A.~G.~Akeroyd, A.~Arhrib and E.~-M.~Naimi,
  Phys.\ Lett.\ B {\bf 490}, 119 (2000)
  [hep-ph/0006035];
  I.~F.~Ginzburg and I.~P.~Ivanov,
  Phys.\ Rev.\ D {\bf 72}, 115010 (2005)
  [hep-ph/0508020].


\bibitem{peskin}
  M.~E.~Peskin and T.~Takeuchi,
  Phys.\ Rev.\ D {\bf 46}, 381 (1992).
  

\bibitem{EWPO-PDG} 
J. Beringer et al. (Particle Data Group), Phys. Rev. D86, 010001 (2012).
  
  
  
\bibitem{Baak:2012kk} 
  M.~Baak, M.~Goebel, J.~Haller, A.~Hoecker, D.~Kennedy, R.~Kogler, K.~Moenig and M.~Schott {\it et al.},
  Eur.\ Phys.\ J.\ C {\bf 72}, 2205 (2012)
  [arXiv:1209.2716 [hep-ph]].
\bibitem{Baak:2013ppa} 
  M.~Baak and R.~Kogler,
  arXiv:1306.0571 [hep-ph].




\bibitem{EWPO-2HDM} 
  H.~-J.~He, N.~Polonsky and S.~-f.~Su,
  Phys.\ Rev.\ D {\bf 64}, 053004 (2001)
  [hep-ph/0102144].

\bibitem{EWPO-2HDM2} 
  S.~Kanemura, Y.~Okada, H.~Taniguchi and K.~Tsumura,
  Phys.\ Lett.\ B {\bf 704}, 303 (2011)
  [arXiv:1108.3297 [hep-ph]].

\bibitem{EWPO-2HDMwScalars1}
W.~Grimus, L.~Lavoura, O.~M.~Ogreid and P.~Osland,
  J.\ Phys.\ G {\bf 35}, 075001 (2008)
  [arXiv:0711.4022 [hep-ph]].
\bibitem{EWPO-2HDMwScalars2}
W.~Grimus, L.~Lavoura, O.~M.~Ogreid and P.~Osland,
  Nucl.\ Phys.\ B {\bf 801}, 81 (2008)
  [arXiv:0802.4353 [hep-ph]].


\bibitem{chargedLEP}
  G.~Abbiendi {\it et al.}  [ALEPH and DELPHI and L3 and OPAL and LEP Collaborations],
  Eur.\ Phys.\ J.\ C {\bf 73}, 2463 (2013)
  [arXiv:1301.6065 [hep-ex]].


\bibitem{chargedLHC}
  G.~Aad {\it et al.}  [ATLAS Collaboration],
  JHEP {\bf 1503} (2015) 088
  [arXiv:1412.6663 [hep-ex]].
CMS Collaboration, CMS-PAS-HIG-14-020, CERN, Geneva Switzerland (2014).



\bibitem{Hermann:2012fc} 
  T.~Hermann, M.~Misiak and M.~Steinhauser,
  JHEP {\bf 1211}, 036 (2012)
  [arXiv:1208.2788 [hep-ph]].


\bibitem{btotaunu}
  I.~Adachi {\it et al.}  [Belle Collaboration],
  Phys.\ Rev.\ Lett.\  {\bf 110}, no. 13, 131801 (2013)
  [arXiv:1208.4678 [hep-ex]].

\bibitem{otherbtotaunu}
  B.~Aubert {\it et al.}  [BaBar Collaboration],
  Phys.\ Rev.\ D {\bf 81}, 051101 (2010)
  [arXiv:0912.2453 [hep-ex]];
  K.~Hara {\it et al.}  [Belle Collaboration],
  Phys.\ Rev.\ D {\bf 82} (2010) 071101
  [arXiv:1006.4201 [hep-ex]].



\bibitem{santos}
  A.~Barroso, P.~M.~Ferreira, R.~Santos, M.~Sher and J.~P.~Silva,
  arXiv:1304.5225 [hep-ph].


\bibitem{pseudoLHC}
  G.~Aad {\it et al.}  [ATLAS Collaboration],
  JHEP {\bf 1302}, 095 (2013)
  [arXiv:1211.6956 [hep-ex]];
CMS Collaboration, CMS-PAS-HIG-13-021.

\bibitem{btodtaunu}
  J.~P.~Lees {\it et al.}  [BaBar Collaboration],
  Phys.\ Rev.\ Lett.\  {\bf 109}, 101802 (2012)
  [arXiv:1205.5442 [hep-ex]].

\bibitem{Ko-B}
  P.~Ko, Y.~Omura and C.~Yu,
  JHEP {\bf 1303}, 151 (2013)
  [arXiv:1212.4607 [hep-ph]].


\bibitem{heavyHiggs}
ATLAS Collaboration, ATLAS-CONF-2013-013, CERN, Geneva Switzerland (2013).



\bibitem{invisible}
  S.~Chatrchyan {\it et al.}  [CMS Collaboration],
  Eur.\ Phys.\ J.\ C {\bf 74} (2014) 2980
  [arXiv:1404.1344 [hep-ex]];
  G.~Aad {\it et al.}  [ATLAS Collaboration],
  Phys.\ Rev.\ Lett.\  {\bf 112} (2014) 201802
  [arXiv:1402.3244 [hep-ex]].

\bibitem{global}
  J.~R.~Espinosa, M.~Muhlleitner, C.~Grojean and M.~Trott,
  JHEP {\bf 1209}, 126 (2012)
  [arXiv:1205.6790 [hep-ph]];
  G.~Belanger, B.~Dumont, U.~Ellwanger, J.~F.~Gunion and S.~Kraml,
  Phys.\ Rev.\ D {\bf 88}, 075008 (2013)
  [arXiv:1306.2941 [hep-ph]];
  S.~Choi, S.~Jung and P.~Ko,
  JHEP {\bf 1310}, 225 (2013)
  [arXiv:1307.3948].

 
 
\bibitem{dijetbound} 
  T.~Aaltonen {\it et al.}  [CDF Collaboration],
  Phys.\ Rev.\ D {\bf 79}, 112002 (2009)
  [arXiv:0812.4036 [hep-ex]].

\bibitem{dijetbound2} 
  S.~Chatrchyan {\it et al.}  [CMS Collaboration],
  Phys.\ Rev.\ D {\bf 87}, 114015 (2013)
  [arXiv:1302.4794 [hep-ex]].

\bibitem{ttbarbound} 
CDF Collaboration, CDF/PUB/EXOTIC/CDFR/10927.
\bibitem{ttbarbound2} 
  G.~Aad {\it et al.}  [ATLAS Collaboration],
  Phys.\ Rev.\ D {\bf 88}, 012004 (2013)
  [arXiv:1305.2756 [hep-ex]].

\bibitem{higgsdiscovery}
  G.~Aad {\it et al.}  [ATLAS Collaboration],
  Phys.\ Lett.\ B {\bf 716}, 1 (2012)
  [arXiv:1207.7214 [hep-ex]];
  S.~Chatrchyan {\it et al.}  [CMS Collaboration],
  Phys.\ Lett.\ B {\bf 716}, 30 (2012)
  [arXiv:1207.7235 [hep-ex]].


\bibitem{hdecay}
 A.~Djouadi, J.~Kalinowski and M.~Spira,
  Comput.\ Phys.\ Commun.\  {\bf 108}, 56 (1998)
  [hep-ph/9704448].

\bibitem{cmsnew}
  V.~Khachatryan {\it et al.}  [CMS Collaboration],
  Eur.\ Phys.\ J.\ C {\bf 74} (2014) 10,  3076
  [arXiv:1407.0558 [hep-ex]];
  S.~Chatrchyan {\it et al.}  [CMS Collaboration],
  Phys.\ Rev.\ D {\bf 89}, 092007 (2014)
  [arXiv:1312.5353 [hep-ex]].

\bibitem{atlasnew}
  G.~Aad {\it et al.}  [ATLAS Collaboration],
  Phys.\ Rev.\ D {\bf 91} (2015) 1,  012006
  [arXiv:1408.5191 [hep-ex]];
  G.~Aad {\it et al.}  [ATLAS Collaboration],
  Phys.\ Rev.\ D {\bf 90} (2014) 11,  112015
  [arXiv:1408.7084 [hep-ex]].

\bibitem{alignment}
  M.~Carena, I.~Low, N.~R.~Shah and C.~E.~M.~Wagner,
  JHEP {\bf 1404}, 015 (2014)
  [arXiv:1310.2248 [hep-ph]];
  P.~S.~B.~Dev and A.~Pilaftsis,
  JHEP {\bf 1412} (2014) 024
  [arXiv:1408.3405 [hep-ph]].

\bibitem{opposite}
  B.~Dumont, J.~F.~Gunion, Y.~Jiang and S.~Kraml,
  Phys.\ Rev.\ D {\bf 90}, 035021 (2014)
  [arXiv:1405.3584 [hep-ph]].

\bibitem{PDG}
K.A. Olive et al. (Particle Data Group), Chin. Phys. C, 38, 090001 (2014).


\bibitem{planck}
  P.~A.~R.~Ade {\it et al.}  [Planck Collaboration],
  Astron.\ Astrophys.\  {\bf 571}, A16 (2014)
  [arXiv:1303.5076 [astro-ph.CO]].



\bibitem{micromegas}
  G.~Belanger, F.~Boudjema, A.~Pukhov and A.~Semenov,
  Comput.\ Phys.\ Commun.\  {\bf 185}, 960 (2014)
  [arXiv:1305.0237 [hep-ph]].


\bibitem{LUX}
  D.~S.~Akerib {\it et al.}  [LUX Collaboration],
  Phys.\ Rev.\ Lett.\  {\bf 112} (2014) 091303
  [arXiv:1310.8214 [astro-ph.CO]].



\bibitem{Ko:2014tca} 
  P.~Ko, Y.~Omura and C.~Yu,
  AIP Conf.\ Proc.\  {\bf 1604}, 210 (2014).

\bibitem{Kniehl:1992ez} 
  B.~A.~Kniehl and H.~-G.~Kohrs,
  Phys.\ Rev.\ D {\bf 48}, 225 (1993).



 


\end{thebibliography}
\end{document}